\documentclass[prl,twocolumn,aps,flushbottom,superscriptaddress,showpacs]{revtex4-1}
\usepackage{xspace,amsmath,amsfonts,amsthm,amssymb,amsbsy,graphicx,color}

\begin{document}
 
\newcommand{\nn}{\nonumber} 
\newcommand{\ms}[1]{\mbox{\scriptsize #1}}
\newcommand{\msi}[1]{\mbox{\scriptsize\textit{#1}}}
\newcommand{\dg}{^\dagger}
\newcommand{\smallfrac}[2]{\mbox{$\frac{#1}{#2}$}}
\newcommand{\la}{\langle}
\newcommand{\ra}{\rangle}
\newcommand{\ket}[1]{| {#1} \ra}
\newcommand{\bra}[1]{\la {#1} |}
\newcommand{\pfpx}[2]{\frac{\partial #1}{\partial #2}}
\newcommand{\dfdx}[2]{\frac{d #1}{d #2}}
\newcommand{\ioh}{-\frac{i}{\hbar}}
\newcommand{\ohh}{-\frac{1}{\hbar^2}}
\newcommand{\half}{\smallfrac{1}{2}}

\newtheorem{theo}{Theorem} 
\newtheorem{lemma}{Lemma}

\title{High-Threshold Low-Overhead Fault-Tolerant Classical Computation and the Replacement of Measurements with Unitary Quantum Gates} 

\author{Benjamin Cruikshank} 
\affiliation{U.S. Army Research Laboratory, Computational and Information Sciences Directorate, Adelphi, Maryland 20783, USA}
\affiliation{Department of Physics, University of Massachusetts at Boston, Boston, MA 02125, USA} 
\author{Kurt Jacobs}
\affiliation{U.S. Army Research Laboratory, Computational and Information Sciences Directorate, Adelphi, Maryland 20783, USA}
\affiliation{Department of Physics, University of Massachusetts at Boston, Boston, MA 02125, USA} 
\affiliation{Hearne Institute for Theoretical Physics, Louisiana State University, Baton Rouge, LA 70803, USA} 

\begin{abstract} 
Von Neumann's classic ``multiplexing'' method is unique in achieving high-threshold fault-tolerant classical computation (FTCC), but has several significant barriers to implementation: i) the extremely complex circuits required by randomized connections, ii) the difficulty of calculating its performance in practical regimes of both code size and logical error rate, and iii) the (perceived) need for large code sizes. Here we present numerical results indicating that the third assertion is false, and introduce a novel scheme that eliminates the two remaining problems while retaining a threshold very close to von Neumann's ideal of $1/6$. We present a simple, highly ordered wiring structure that vastly reduces the circuit complexity, demonstrates that randomization is unnecessary, and provides a feasible method to calculate the performance. This in turn allows us to show that the scheme requires only moderate code sizes, vastly outperforms concatenation schemes, and under a standard error model a unitary implementation realizes universal FTCC with an accuracy threshold of $p < 5.5\%$, in which $p$ is the error probability for 3-qubit gates. FTCC is a key component in realizing measurement-free protocols for quantum information processing. In view of this we use our scheme to show that all-unitary quantum circuits can reproduce any measurement-based feedback process in which the asymptotic error probabilities for the measurement and feedback are $(32/63) p \approx 0.51 p$ and $1.51 p$, respectively.   
\end{abstract} 

\maketitle  

The problem of performing classical computing reliably with unreliable logic gates is referred to as fault-tolerant classical computation (FTCC). The first method for realizing FTCC was devised by von Neumann, who called it \textit{multiplexing}~\cite{vonNeumann56}. It achieves what may be the highest possible error threshold (the maximum stable component-wise error rate), but has hitherto been viewed as impracticable. This is due to an apparent need for high redundancy (number of fundamental components required to construct a noise-free logical gate), the need to continually connect and reconnect bits ``€œat random''€ at a potentially large spatial separation, and the difficulty of both analytically calculating the performance for moderate code sizes and simulating the performance in the low-error regimes required for reliable computation ~\cite{vonNeumann56, Pippenger90, Nikolic02, Roy05, Bhaduri07, Beiu07, Ma08, Han11, Sen16}. The field of probabilistic cellular automata was partially motivated by addressing the second problem, but has not to-date produced a complete and feasible FTCC scheme~\cite{Toom80,Gacs83, Gacs85, Han03, Peper04, Di08, Zloudek11, deMaere12, Ponselet13, Mairesse14, Slowinski15, Lee16}. A second method for FTCC was developed more recently in the context of quantum computing, and involves ``concatenating'' error-correction codes and logic gates~\cite{Shor96, Kitaev97, Aharonov98, Knill98, Preskill98, Gottesman98, Steane99, Knill05, Terhal05, Nielsen05, Aliferis06, Szkopek06, Aliferis06b, Dawson06, Svore07, Aharonov08, Aliferis08, Fujii10, Koenig10, Paetznick13, Jones13, Jochym14, Stephens14, Campbell14, Bombin15}. However, the concrete FTCC schemes developed using concatenation require high redundancy and connections between widely separated code bits, and have not to-date achieved the high thresholds of multiplexing schemes~\cite{Boykin05, Antonio16x}. 

We are interested in FTCC here primarily because of its central role in the question of the importance of measurements in realizing physical processes and control protocols. From a fundamental point of view, measurements play no special role in physical processes: all dynamics generated by measurement and feedback processes (including those involving post-selection~\footnote{Calculations of the error rates for the gates, error-correction circuits, and universal computation, as well as a brief discussion of postselection, are given in the supplemental material XXXXX, which includes Refs.~\cite{Lee08, Hoe10}}) can be reproduced by unitary evolution~\cite{Jacobs14}. Consequently the utility of measurements in any physical protocol arises only from technological constraints. For fault-tolerant quantum computation (FTQC), in which all high-threshold schemes to-date employ measurements~\cite{Knill05, Aliferis08, Fowler09, Stephens14}, the constraint is a fixed error-probability $p$ for all quantum gates and measurements. 

Interest in measurement-free (or measurement-light) quantum computing~\cite{DiVincenzo07, Paz10, Fitz09, Fujii14, Herold15} is motivated by the fact that implementing large numbers of measurements on many qubits requires an additional technological overhead beyond that of unitary circuits. To this end schemes have been devised in which measurements can be noisy and/or slow~\cite{DiVincenzo07, Paz10}, and quite recently automata-based methods were introduced for eliminating both measurements and the high processing overhead involved in correcting surface codes~\cite{Fujii14, Herold15}. Here our purpose is to determine the ability of unitary circuits to perform the functions of arbitrary measurement procedures. 

For the purposes of FTQC (or any quantum process subjected to errors) the physical addition provided by measurements is \textit{amplification}: measurements are defined as producing a result that can be processed on a classical digital computer. The resulting error-free classical processing is the sole advantage of measurements. As a result the question of the importance of measurement to quantum processes is intimately related to how well mesoscopic gates (those with error $p$) can perform such error-free classical processing, and thus to the fundamental limits of FTCC. 

Our first main result is an explicit scheme for FTCC with unitary gates that largely solves the long-standing problems with von Neumann's multiplexing method, while achieving almost the same high threshold.  We then apply our FTCC protocol to the problem of using unitary circuits to reproduce a general measurement and feedback process. Our second main result is that, with a threshold of $p = 2.8\%$, unitary circuits can do this and achieve effective measurement and feedback errors close to $p$. It should be noted that this result does not imply alone that a given measurement-based protocol can be replaced by a unitary one; additional technological factors, such as the time taken by the processing circuits, may also play an important role in an implementation. This result does show that unitary circuits can effectively realize high-fidelity measurements, and that perfect classical processing within quantum mesoscopic circuits is entirely feasible. They also furnish a new tool for understanding the power of unitary protocols.

Before we present our error-correction scheme, it is worth discussing the key issues with von Neumann's method in more detail. Von Neumann's scheme uses a repetition code, and corrects errors in the code by applying a ``majority counting'' gate to triples of code bits. The output of this gate is a single bit whose value is that of the majority of the three input bits. The output bit is then copied to produce three bits that are the corrected versions of the input bits. As noted above, one of the primary problems with the scheme is the need to reduce correlated errors by randomly selecting triples across multiple repetitions of the error-correction. This virtually prohibits the use of fixed wiring for the gate interconnections due to the complexity of the resulting circuits. Furthermore, an appeal to truly ``random" fixed connections in a computational circuit implies the need for unique random reconnections at every correction step of a computation of any length, something that is clearly infeasible. A possible solution is to dynamically reconnect the gates at each correction step. While such a process cannot be used in a fundamental theory of fault-tolerance (the logic circuits that generate and store the new connections will introduce further errors), it could be used to  implement FTCC with mesoscopic circuits by using error-free macroscopic classical computers to perform the dynamical reconnection. Nevertheless, doing so requires the significant additional overhead of complex classical control circuits that allow any triple of code bits to interact at any time. Our solution completes the theory of multiplexing by eliminating randomness altogether and allowing error correction to be performed by a small set of fixed connections within the code, which are recycled over a short sequence. We note that our scheme has a single restriction over von Neumann's, which is that the code size must be a power of 3. 

A further advantage we provide over randomized multiplexing is that with multiplexing the logical error rate for moderate, realistic code sizes can be obtained only by numerical simulation. This is problematic because i) a simulation must generate at least tens of logical failures to obtain any accuracy, and ii) for realistic computing applications the error probability per logical bit must be very small (e.g. $10^{-12}$). As a result the problem would be challenging even for modern large-scale parallel machines.  

\begin{figure}[t]
\leavevmode\includegraphics[width=0.85\hsize]{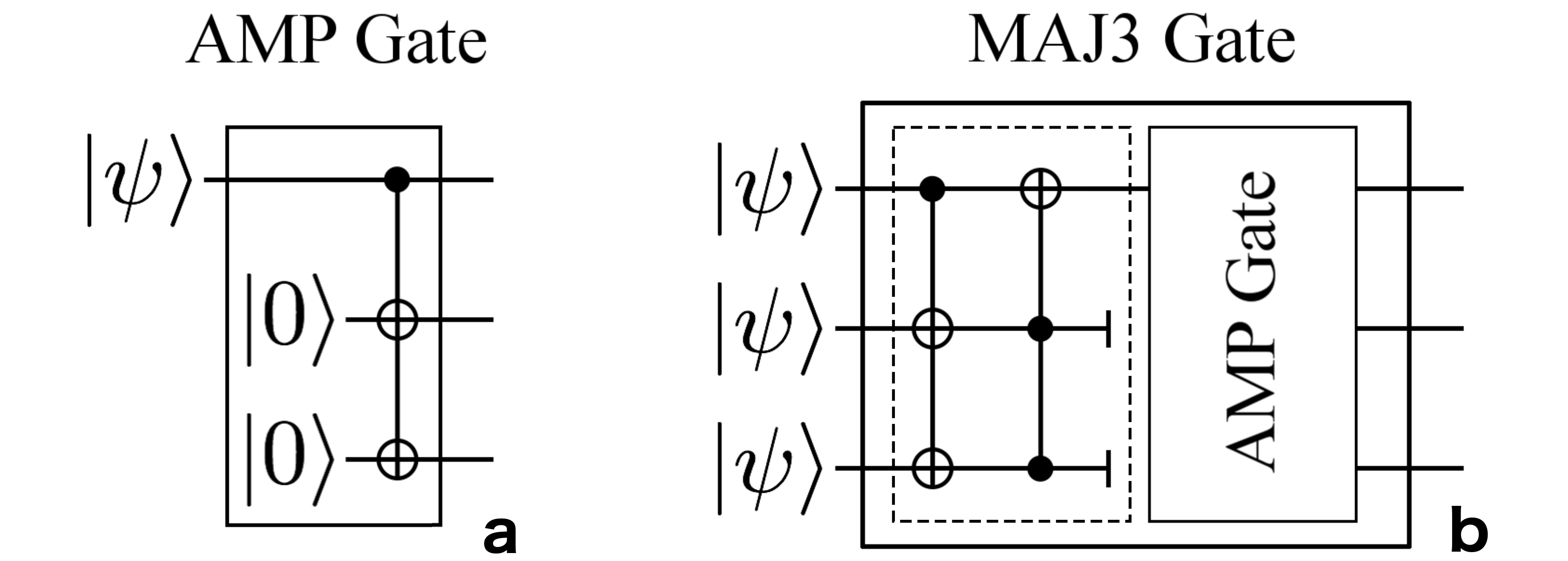}
\caption{Unitary constructions for the (a) AMP and (b) MAJ3 gates defined in the text. The unitary version of AMP copies the input qubit in the computational basis by applying a controlled-not (CNOT) gate from this input to each of two qubits prepared in the 0 state. The dashed box is the MAJ1 gate.} 
\label{fig1} 
\end{figure}

We now describe our FTCC scheme, beginning with the logic gates from which it is built: 

\textit{Elementary 3-bit gates:} We define a MAJ1 gate (the name deriving from ``majority'') as von Neumann's 3-bit majority counting gate, described above.  We define a gate AMP (short for ``amplification'') that takes in one bit and outputs three copies of it. Finally, we define a gate MAJ3 as a MAJ1 that has three outputs, being the usual MAJ1 output bit and two additional copies of it. Thus the MAJ3 is a MAJ1 followed by an AMP. We must be able to implement all these gates unitarily, so we present explicit unitary versions of them in Fig.~\ref{fig1}. These unitary versions are shown in terms of CNOT and Toffoli gates~\cite{mikeandike}, but our error model treats the MAJ1 and AMP as elementary 3-bit unitary processes. 

\textit{Error-correcting scheme:} The logical values of bits on which we wish to do computation are stored in a simple repetition code of size $3^{n+1}$ where n is a non negative integer called the \textit{level} of the code. These bits can be arranged in a hypercube of dimension $n+1$ with side length 3. Error correction is now achieved by applying $n$ parallel MAJ3 gates along each of the $n+1$ axes in sequence. This implies that the logical value can be equivalently thought of as stored in a network of $3^n$ MAJ3 gates that interact sequentially along each axis of an n-dimensional hypercube~\cite{Note1}. 

\textit{Performance of the error-correcting circuit:} Our error model is defined by assigning, in the standard way~\cite{Knill05, Fowler12}, a total intrinsic error probability, $p$, to each of the 3-bit unitary quantum gates MAJ1 and AMP. To determine the performance of the error-correction circuit we need the probability, $\varepsilon$, that there is an error in at least one of the output lines of the MAJ3. Given the quantum error model, along with an additional error of $(2/3)p$ in each output to account for errors in the connecting wires and reset locations, we obtain a strict overestimate for $\varepsilon$ to be $(52/21)p \approx 2.3 p$~\footnote{It should be noted that $\varepsilon$ (along with its threshold) is more fundamental to the MAJ3 network than $p$. The former is an error rate for any module which acts as a MAJ3. The latter depends on our unitary implementation of the MAJ3 and our quantum error model~\cite{Note1}}.) We need to obtain the steady-state logical error probability, $p_{\ms{ss}}$, that is maintained by repeated applications of the correction circuit~\footnote{Decoding the logical state (reading it out) is not required for our application here, but we treat it in the supplement~\cite{Note1}.}. Calculating $p_{\ms{ss}}$ is non-trivial, however: straightforward simulations are impractical as discuss above.  We are able to calculate $p_{\ms{ss}}$ for $n=2$ and $n=3$ (codes with 27 and 81 bits, respectively) by mapping the error dynamics to a jump process that has only 3 effective states for $n=2$ and 7 for $n=3$ (details are given in the supplemental material). In Fig.~\ref{fig2} we plot $p_{\ms{ss}}$ as a function of $\varepsilon$ for $n=2$ and $3$. We see from these plots that the error decreases doubly-exponentially with $n$ for $n = 2$ and $3$, and so we expect this to continue for larger values of $n$. This shows a (gatewise) redundancy which scales as $3^n$. In comparison, the redundancy for typical concatenation schemes is $21^n$~\cite{Boykin05, Antonio16x}. We note that for $p = 0.4\%$ and $n=3$ one has $p_{\ms{ss}} = 1.5\times 10^{-11}$, thus small values of $n$ would likely suffice for applications. We obtain a lower bound on the threshold for $n=3$, shown in Fig.~\ref{fig3}a, to be $\varepsilon \approx 15\%$, very close to von Neumann's value of $1/6$. 

\textit{Performance of von Neumann's multiplexing:} In the inset in Fig.\ref{fig2} we compare the performance of von Neumann's scheme to ours for a code size of 81 bits. As noted above this calculation is limited to relatively large error rates. We see that the performance of von Neumann's scheme oscillates about ours, and thus achieves similar performance at high logical error rates. This provides support for the conjecture that a randomized scheme should perform at least as well as ours. Nevertheless, one cannot merely extrapolate to small error rates, as made especially clear by the oscillatory behavior. 

\begin{figure}[t]
\leavevmode\includegraphics[width=1\hsize]{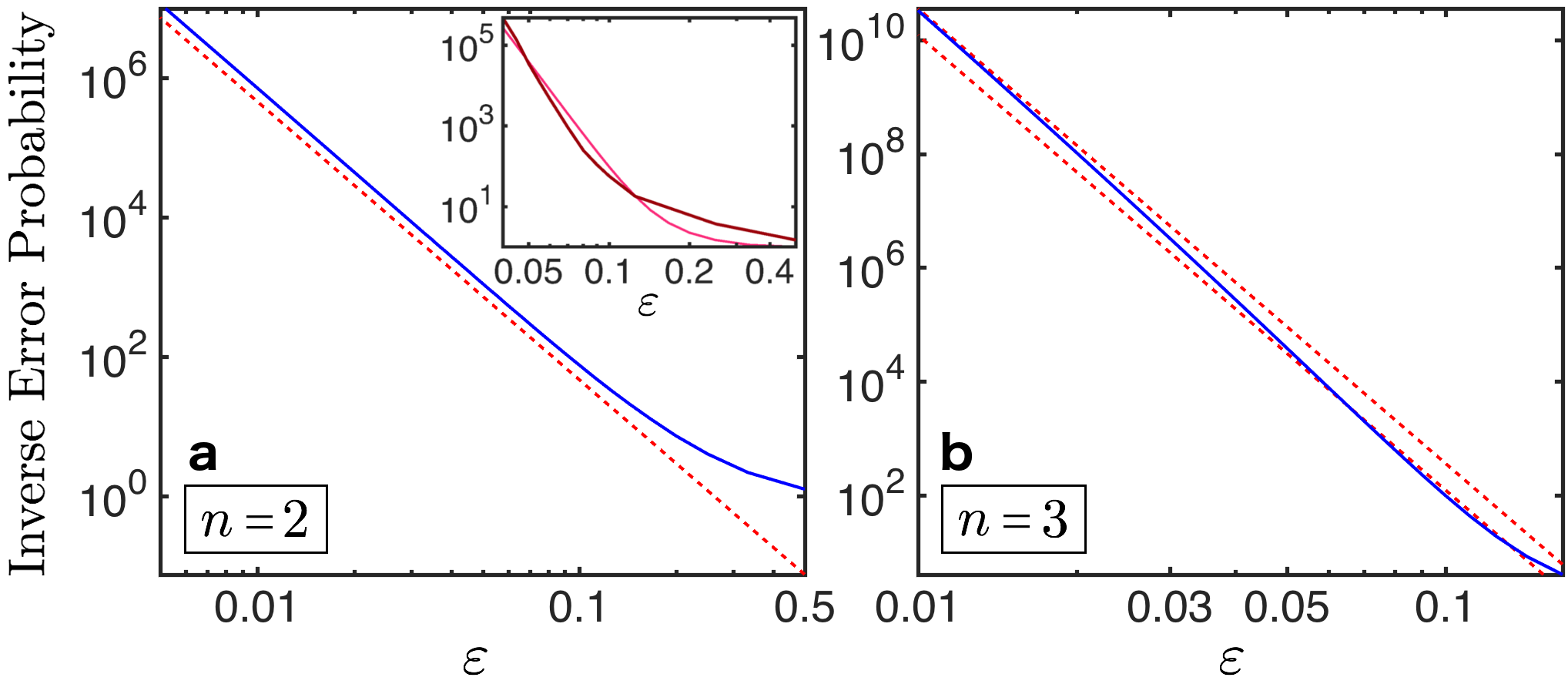}
\caption{(Color online) (a) Solid line: the inverse of the logical error probability, $p_{\ms{ss}}$, for our error correction circuit with a 27-bit code. Dashed line: $1/p_{\ms{ss}}$ for a hypothetical concatenation scheme with a threshold of $1/6$ \cite{mikeandike, Note1}. (b) Solid line: $1/p_{\ms{ss}}$ for our scheme with an 81-bit code. Dashed lines: $1/p_{\ms{ss}}$ for hypothetical concatenation schemes with thresholds of $1/6$ (upper line),  and $1/7$ (lower line). (No such concatenation schemes are known to-date.) The inset shows the performance of von Neumann's scheme (dark line) against the new scheme (light line) for 81 bits, for a range of $\varepsilon$ in which it is feasible to simulate.} 
\label{fig2} 
\end{figure}

\textit{Threshold for universal computation:} Reliable MAJ1 and AMP gates can be used for universal computation. A MAJ1 gate can be used as an AND or OR gate. An AMP gate is composed of CNOTs, which can be used as NOT gates and/or simulated line splitting. (Details can be found in~\cite{Note1, vonNeumann56}.) The most complex construction in our scheme is a coded MAJ1 gate, which consists of a transversal application of MAJ1 gates on three coded bits. The bitwise error rate in an error correcting network at threshold is 1/2. Since the output of a three-input computational gate is necessarily noisier than any one of the inputs, we must have input errors less than 1/2, so the component-wise threshold for universal computation must smaller than 1/6. We recognize the threshold as the basic error rate at which error rates in the outputs of the MAJ1 gates are equal to 1/2. Taking into account the steady-state bitwise error rate of our coded input bits, we find the threshold for universal computation to be $p=5.5\%$, or $\epsilon\approx12.7\%$\cite{Note1}. 

\textit{Scaling of wire length with code size:} A crucial issue for fault-tolerant computation is how the wire length required by the correction and computation circuits increases with code size. Since it is reasonable to suggest that fundamental error rates will increase exponentially with the wire length (the distance between interacting gates/bits), an error correction scheme must be compatible with a compact wiring. Part of our solution is the observation, noted above, that using our method code sizes of no more than $n=4$ (and likely $n=3$) can be expected to be sufficient for any application. If we reroute the wiring from the gate outputs to the inputs, the full error correction circuit for $n=3$ can be executed with a cube of 27 MAJ3 gates. The rerouting need only span the cube separately in each direction, giving a wire length of $2$ (in units of the distance between adjacent gates). The transversal AND gate for logic between coded bits, when we place three code cubes in a row, requires wires of length $3$. For $n=4$ there are also very efficient arrangements. For a single coded bit we now require three cubes of 27 MAJ3 gates for error correction. In Fig.~\ref{fig3}b we show that, arranging these three cubes in an ``L'' configuration, the longest wires required for universal computation have length $3\sqrt{2}$. Only a moderate increase in overall distance is therefore required to implement FTCC.  One could alternatively use an array of static qubits rather than static gates. Under this architecture the error-correction circuits are similarly compact; each qubit need interact with only $2(n+1)$ others~\footnote{For $n=3$ with the qubits arranged in a square, the maximum number of qubits that lie between any two that must interact is 5, always along straight lines in the array. If three dimensions are utilized, this same maximum distance still applies for universal computation at $n=4$, which can be accomplished with a $9\times9\times9$ array.}. 
 
\begin{figure}[t]
\leavevmode\includegraphics[width=1\hsize]{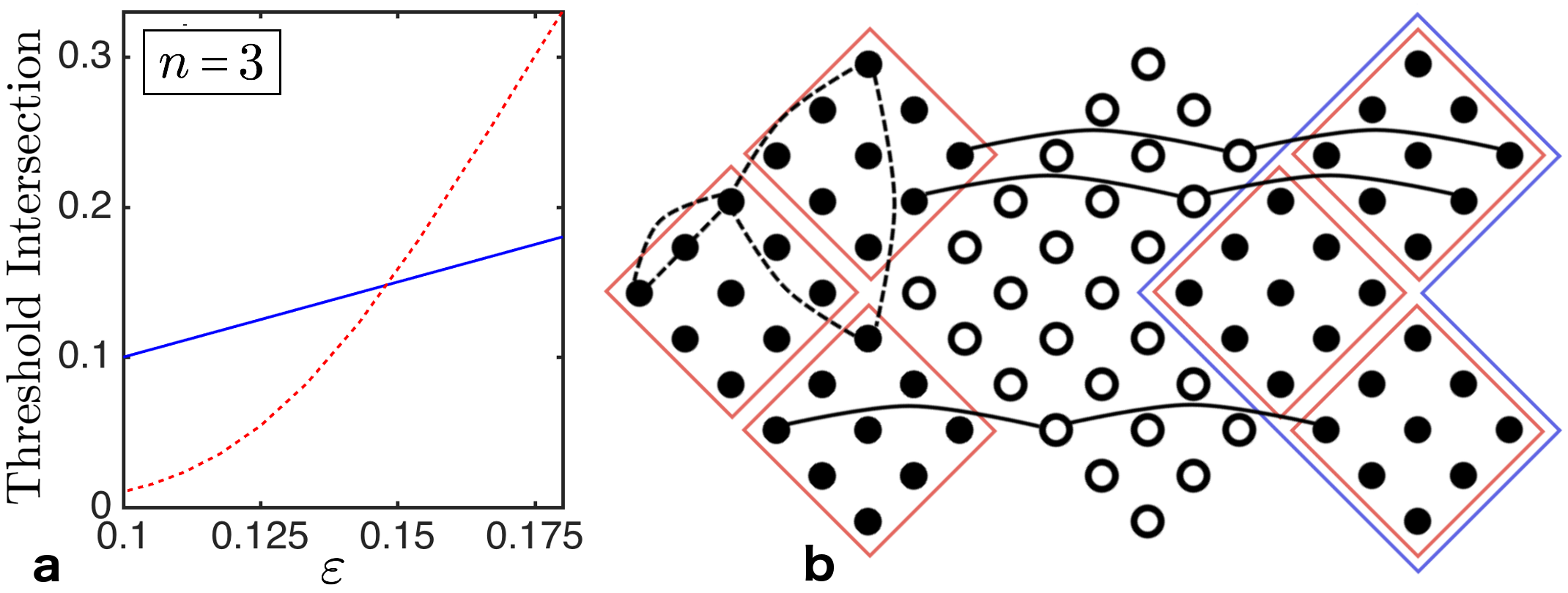}
\caption{(a) The solid line is $\varepsilon$ and the dashed line is the resulting error probability for a code with $n=3$. The point at which these lines cross gives the threshold. (b) Here we show the wiring lengths required for a code with $n=4$ (243 bits). Each dot and circle represents a MAJ3 gate, so that the squares indicate a cube of $27$ MAJ3 gates (an $n=3$, $81$-bit code) viewed from the top. The blue line encircles three of these cubes that form a single $n=4$ code block. The curved solid lines are examples of wires required to implement an AND gate on the two code blocks consisting of black dots. The dashed lines are examples of wires required to implement error correction on an individual code block.} 
\label{fig3} 
\end{figure} 

\textit{Replacing measurements with unitary circuits:} The scheme for efficient FTCC presented above allows us to obtain unitary circuits that perform the role of measurement and feedback processes, and do so with very low error rates. First, we note that the role of every measurement in any physical protocol involves no more than i) classical processing of the measurement result, and ii) at some point in the process, the use of the processed result to apply an operation to a quantum system. We can assume the feedback operation is unitary without loss of generality. We will also assume, without significant loss of generality, that the quantum systems involved are qubits. 

To reproduce the action of a ``black box'' that implements measurement and feedback our unitary circuit must i) encode the qubits to be measured (the input qubits) so that the classical information they contain can be processed by our FTCC protocol, ii) perform the processing, iii) use the processed information (the output qubits), which is stored as a repetition code, to apply a unitary gate to one or more ``target'' qubits. Steps i) and iii) correspond to the processes of measurement and feedback, respectively. In step i), any error in the encoded logical bits introduced by the encoding procedure is precisely equivalent to the measurement error. For step iii), since we must use the state of a single qubit as a control for the feedback operation, any error by which this qubit deviates from the encoded logical output bit is simply an additional probability of error over that of the feedback applied by a classical device. 

To encode the information stored in the computational basis of a qubit we use a circuit consisting of AMP gates. An AMP gate is used to make three copies of the initial qubit in computational basis, and then each of these is tripled again by feeding it into an AMP gate. By repeating this process we produce $3^{n+1}$ qubits that constitute the bits of our repetition code. The key quantity of interest is the probability, $p_{\ms{enc}}$, that the resulting code fails to correctly reflect the state of the bit contained by the initial qubit (more precisely, the probability that the code bits are left in a joint state that will fail to be properly corrected by the error-correction circuits). Calculating $p_{\ms{enc}}$ is a complex task, since we must take into account the correlations formed between the code bits/qubits during the encoding, as well as the action of our error-correction procedure. We obtain, for $n=3$, a strict over-estimate of the encoding error as a $9^{\msi{th}}$-order polynomial in $p$, the full expression for which is given in the supplemental material. The most important property of $p_{\ms{enc}}$ is $p_{\ms{crit}}$, defined as the value of $p$ for which $p_{\ms{enc}} = p$, and for which $p_{\ms{enc}} < p$ whenever $p < p_{\ms{crit}}$. The encoding circuit for $n=3$ has $p_{\ms{crit}} = 2.8\%$, and when $p \ll 2.8\%$ the relationship is $p_{\ms{enc}} \approx (32/63)p \approx 0.51 p$. This encoding error is precisely the measurement error of the black box being simulated. 

Once the classical information in the qubits input to our unitary circuit has been encoded, it can be processed essentially error-free using the error-correction and computation methods present above. Thus it remains to apply an operation to $m$ qubits that is conditional on the processed information. The bits containing this information are stored in our repetition code. To ensure that the feedback correctly mimics the operation of feedback applied by a classical controller we must take account of the following: i) A classical controller is assumed to be error-free, and so does not introduce errors that are correlated between the $m$ target bits; 2) the feedback operation must be implemented with a single control qubit for each target qubit because we are restricted to mesoscopic circuits. Fortunately we can satisfy both demands. Transversal CNOT operations can copy logical bits fault-tolerantly. This can provide an ensemble of $m$ logical bits that are essentially correct (to the basic logical failure rate). Applying a series of MAJ1 gates to each logical bit (using $n+1$ iterations for a code at level $n$), we can provide $m$ qubits with independent errors which approach (to first order in $p$) the individual failure rate of the MAJ1~\cite{Paz10}. Using these qubits as the controls for the feedback operations gives an error of $(32/63)p \approx 0.51 p$ over that of the classical feedback operation. Such an additional error in the feedback appears to be a necessary consequence of the use of mesoscopic circuits for this purpose. 

Here we have presented a scheme for fault-tolerant classical computation that significantly outperforms all previous schemes. In doing so we have shown that multiplexing neither requires randomization, nor large code sizes as have previously been thought. We have used this new scheme to show that unitary mesoscopic circuits can perform all functions of measurement with errors that remain very close to $p$. 

\textit{Acknowledgments:} This research was supported in part by the NSF project PHY-1212413 and by an appointment to the Student Research Participation Program at the U.S Army Research Laboratory administered by the Oak Ridge Institute for Science and Education through an interagency agreement between the U.S Department of Energy and USARL.

\begin{center}
\textbf{\large Supplemental Materials: High-Threshold Low-Overhead Fault-Tolerant Classical Computation and the Replacement of Measurements with Unitary Quantum Gates}
\end{center}
\setcounter{equation}{0}
\setcounter{figure}{0}
\setcounter{table}{0}
\makeatletter
\renewcommand{\theequation}{S\arabic{equation}}
\renewcommand{\thefigure}{S\arabic{figure}}





\subsection{Introduction} 
In this supplemental material we give the details of how we analyze the error dynamics and calculate the performance of i) the fault-tolerant classical error-correction method that we present in the paper, and ii) the encoding process that is an essential part of the method we present to replace measurements with unitary circuits. We also discuss some further topics that may be of interest to readers unfamiliar with them: i) we summarize prior research on classical fault tolerant error correction and computation, including the original multiplexing scheme by von Neumann; ii) we discuss how post-selection processes are replaced by fixed unitary circuits.

\subsection{Background: Multiplexing and other methods}
\label{secFTCC}

We begin by describing John von Neumann's classic multiplexing method \cite{vonNeumann56} as this is the starting point for our scheme. Consider that we have a classical bit encoded in three bits so that 000 codes for 0 and 111 codes for 1. This is a repetition code and is easily generalized to any number of bits. John von Neumann (JvN) defined a gate with three inputs and three outputs that corrects any single error for the 3-bit code, and called it the "majority organ''. We will call it the MAJ3 gate. Because all our gates necessarily have errors, the MAJ3 gate is given a probability $\varepsilon$ of failing. Specifically, if $X$ denotes the state of a single bit, and $\bar{X}$ denotes $\mbox{NOT}(X)$, then given any of the four inputs $XXX$, $\bar{X}XX$, $X\bar{X}X$, or $XX\bar{X}$, the MAJ3 gate produces the output $XXX$ with probability $(1-\varepsilon)$ and the output $\bar{X}\bar{X}\bar{X}$ with probability $\varepsilon$.  

In JvN's scheme a repetition code using $N$ bits, in which $N$ is a multiple of three, is fed into $N/3$ MAJ3 gates to perform a correction operation, called a \textit{restorative phase}. The outputs of these gates are rearranged in a random way and then fed into another set of MAJ3 gates to perform another restorative phase. This sequence of restorative phases is repeated as many times as desired. JvN showed that this circuit performs fault-tolerant error correction on the repetition code. The first step of his analysis is to consider a single MAJ3 gate and to show that if the inputs have \textit{independent errors} then there is a stable error probability $\eta$ for the input bits that will be preserved by the gate. This is not difficult to show, and so long as $\varepsilon \leq 1/6$ the stable error probability is 
\begin{align}
\eta = \frac{1}{2} \left( 1 \pm \sqrt{\frac{1-6\varepsilon}{1-2\varepsilon}} \right) .  
   \label{pss1}
\end{align} 
The two stable solutions, which for $\varepsilon \ll 1$ are $\eta \approx \varepsilon$ and $\eta \approx 1- \varepsilon$, correspond to the two logical states we wish to encode in the network. 

We must now ensure that the inputs to the MAJ3 gates in each restorative phase are independent, even though the outputs from the MAJ3 gates from the previous restorative phase are not. This is the second part of JvN's approach, and is the purpose of randomizing the connections from the outputs of one restorative phase of MAJ3 gates to the inputs of those of the next restorative phase. If $N$ is large then the randomization means that for each MAJ3 gate the probability that more than one of its inputs comes from the outputs of the same MAJ3 gate is small. This probability goes down as $N$ increases, so for sufficiently large $N$ the inputs are independent to good approximation. 

People have studied Von Neumann's concept under different basic gate architectures,  e.g.\ being built from NAND gates \cite{Ma08, Roy05, Bhaduri07}, having larger majority votes than best-of-three \cite{Beiu07}, and finding an optimal number and placement of restorative phases in certain circuits \cite{Ma08, Roy05, Bhaduri07}. See \cite{Han11} for a contemporary review, and for a review of the initial research see \cite{Pippenger90}. Von Neumann's method is still pointed too as a potential architecture for networks of nanocircuits \cite{Sen16}. One major impracticality, however, of the multiplexing concept is the ``random'' rewiring required at each restorative phase. This implies two major difficulties: The first is an unfortunate choice between either a real-time random rewiring of the system at each restorative phase, or a number of static ``random'' rewirings which is equal to the number of required restorative phases. The second difficulty is another unavoidable technical complication which is inherent in randomly reconnecting gates which comprise a large repetition code in physical space. If gates are allowed to interact with any given gate in the code, we must allow for wires which can carry a signal farther than the typical ``nearest neighbors'' interaction, and any wire must be expected to have a signal degradation which is exponential in its length \cite{Gacs83,Pippenger90}. Another impracticality is that the code size required for these random constructions, as well as explicit constructions based on randomization, has been estimated to be large \cite{vonNeumann56, Pippenger90, Nikolic02, Peper04}. 

The aversion to long wires spawned the study of universal computation by arrays with noisy nearest neighbor interactions, called \textit{probabilistic cellular automata}. The first proofs that this was possible were given by G{\'a}cs \cite{Gacs85, Gacs83}, following the work of Toom \cite{Toom80}. The challenges to practicality inherent in the field are that cells themselves generally require a rather complex functionality, and the initial estimate for thresholds of fault-tolerance were rather low \cite{Gacs85}. Due to the complexity of the topic, recent authors have presented schemes which are only partially fault-tolerant. Though approaches to fault-tolerance vary, some of the componentry in the scheme is assumed to work perfectly \cite{Lee08, Lee16}, often the error correcting circuitry itself \cite{Han03, Peper04}. For example, the faults are often considered to be permanent \textit{defects} which must be replaced or worked around \cite{Di08, Nikolic02, Zloudek11}, as opposed to unavoidable \textit{transient} faults in working components, as in the quantum model. In \cite{Hoe10}, the author has returned to allowing for interactions with ``far" neighbors. The quantification of the behavior of probabilistic cellular automata continues to be an area of research \cite{deMaere12, Ponselet13, Mairesse14,Slowinski15}, and the problem of a practical fault-tolerant computation scheme appears to be open. 

We address these difficulties by eliminating the randomness in the wiring of the restorative phase of JvN's multiplexing scheme. We develop a simple systematic wiring scheme for MAJ3 gates which is motivated by the concatenation structure common in Fault Tolerant Quantum Computation. It uses reversible (unitary) gates which makes it suitable for quantum applications. For a lower-threshold scheme with similar motivation and fundamental gates, see \cite{Boykin05}. We explicitly calculate the failure rate of an error correcting network which comprises twenty seven gate locations which are subject to a simple sequence of interactions (wirings). The length of the wires, explicitly detailed, is minimized by the small size of the code and the geometry of the actual scheme. Furthermore, the systematic wiring scheme has close to the ideal Von Neumann error-correction threshold of $\epsilon = 1/6$, and has a network error rate which decays super exponentially in the gate overhead, and thus achieves reliable computation at low gate numbers. The method can be extended to a higher code size if further error suppression is required, at a cost of a small increase in wire length. 

\subsection{Performance of the error-correction scheme}
\label{secRestorativePhase}

Here we introduce the level-1 MAJ gate, constructed from von Neumann's 3-input MAJ3 gate. We then introduce the error-correction scheme for code sizes $n=1$, $2$, and $3$, by building each of the higher levels from the previous level. We then calculate explicitly the performance for $n=2$ and $3$ in terms of $\varepsilon$, which is the basic probability of error of the 3-input MAJ3 gate. 

Without error correction, a bit stored in a noisy physical implementation is limited in its usefulness by its fundamental error rate. Even a repetition code of an infinite number of these bits becomes useless in protecting the logical state for durations over which the probability of an error in a given noisy bit approaches $1/2$. The purpose of a good error correcting network is to suppress the decay rate of the encoded logical state to arbitrarily low levels, so that state storage can be attained with high confidence for any length of time. Furthermore, the size of the error-correction circuit (the ``redundancy'' in the network) should be a manageable function of the suppressed error rate. In addition, the fundamental bit-wise error rate that the correction network is able to tolerate, called the \textit{threshold}, should be as high as possible. 

\subsubsection{The level-1 MAJ gate} 

\begin{figure}[b]
\leavevmode\includegraphics[width=1\hsize]{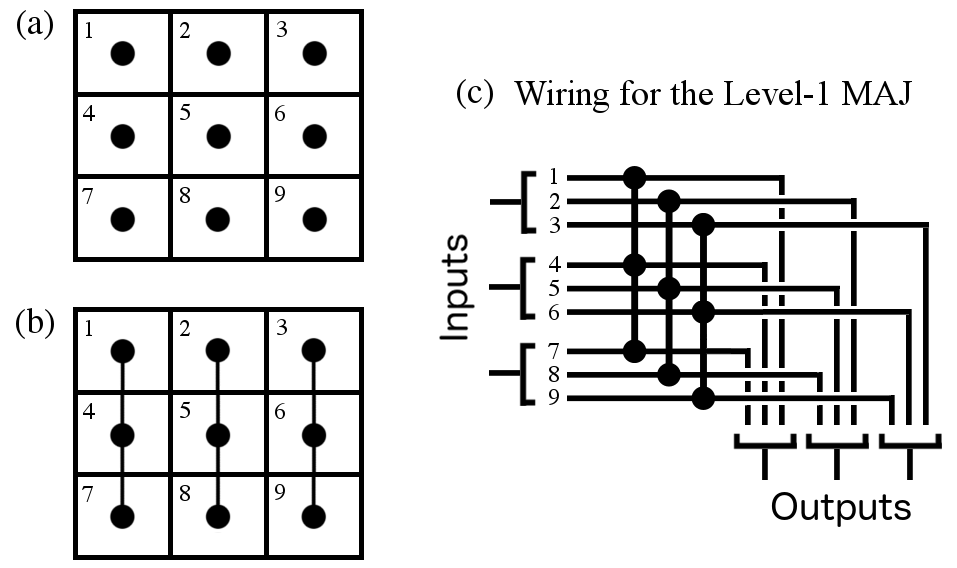}
\caption{In (a) we depict nine bits/qubits arranged in a 3-by-3 square. In (b) the lines that connect the triples of qubits in each column represent the application of a MAJ3 gate to each triple. To obtain good error correction for the nine qubits we must alternate applying MAJ3 gates to the triples that form the rows and those that form the columns. We therefore define a \textit{level-1 MAJ gate} as depicted in (c). Each of the three inputs to the level-1 MAJ gate is a triple of qubits. The outputs of the three MAJ3 gates that make up the level-1 MAJ gate are re-routed so if the three inputs to the level-1 MAJ are the rows of the square, then the three outputs are the columns of the square. We can then apply a sequence of level-1 MAJ gates to error-correct the nine qubits.} 
\label{figS1} 
\end{figure}

We first note that running a series of MAJ3 gates sequentially on the same three bits can only have a detrimental effect on the reliability of the logical (encoded) state. This is because the output of the MAJ3 has the error rate $\varepsilon$, which is greater than that of an individual bit, and, further, it is correlated among the three bits. Nevertheless a suitably designed network of MAJ3 gates can significantly improve the reliability of the encoded state.  Consider the application of 3 MAJ3's on 9 bits arranged in a square matrix as depicted in Fig.~\ref{figS1}. 
We will call this configuration of MAJ3 gates a \textit{level-1 MAJ gate}. We can think of the inputs to the level-1 MAJ gate as being the \textit{rows} of the matrix, and think of the outputs as being the \textit{columns}. If two of the input rows are in error simultaneously, the output of the level-1 MAJ will be completely in error. Otherwise, the error probabilities of each of the output columns are mutually independent. In this analysis, we will make the major simplification of ignoring the cases in which there are simultaneously multiple input errors and errors in the MAJ3 gates, so that the latter happen to correct the former. That is, we obtain an over-estimate of the total errors by ignoring the beneficial probability that our gates can ``accidentally'' correct a bad input state. 

\begin{figure}[b]
\leavevmode\includegraphics[width=0.98\hsize]{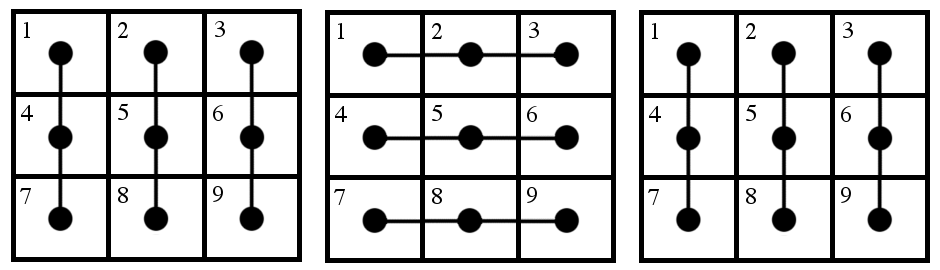}
\caption{Here we depict a sequence of applications of MAJ3 gates to 9 qubits arranged in a 3-by-3 square so as to perform error-correction on a 9-bit repetition code. Each of lines connecting three qubits with a square represents the application of a MAJ3 gate to those qubits. Reading left to right, first three MAJ3 gates are applied to the columns of the square, then three MAJ3's are applied to the rows, and finally again to the columns. This sequence of MAJ3's is equivalent to a sequence of three level-1 MAJ gates applied to the 9 qubits. The level-1 MAJ gate is defined in Fig.~\ref{figS1}.} 
\label{figS2} 
\end{figure}

Note that the fact we take the \textit{rows} of the matrix of nine bits as the inputs, and the \textit{columns} of the matrix as the outputs is crucial in obtaining the necessary error-correction properties of the level-1 MAJ. With this identification we can wire the three outputs (the columns) of a level-1 MAJ into the three inputs (the rows) of another level-1 MAJ, and continue this process by wiring the resulting outputs to the inputs of a third level-1 MAJ, etc. In Fig.~\ref{figS2} we show three level-1 MAJ gates applied sequentially in this way. 
Since single input errors in the MAJ3 have no effect on the output, an entire row that is in error has no effect on the output of the level-1 MAJ. From this we can see immediately that the wiring scheme in Fig.~\ref{figS2} has exactly the same failure properties as a three bit repetition code. That is, for every error-correction step equal in duration to that of a level-1 MAJ (which is also that of a MAJ3), at least two MAJ3 gates must fail simultaneously to incur a logical error. So long as $\varepsilon$ is less than $1/2$, this scheme will offer some improvement on the decay rate of the logical state, which is then no larger than $3\varepsilon^2$.

\subsubsection{The correction circuit for $n=2$} 

\begin{figure}[b]
\leavevmode\includegraphics[width=0.98\hsize]{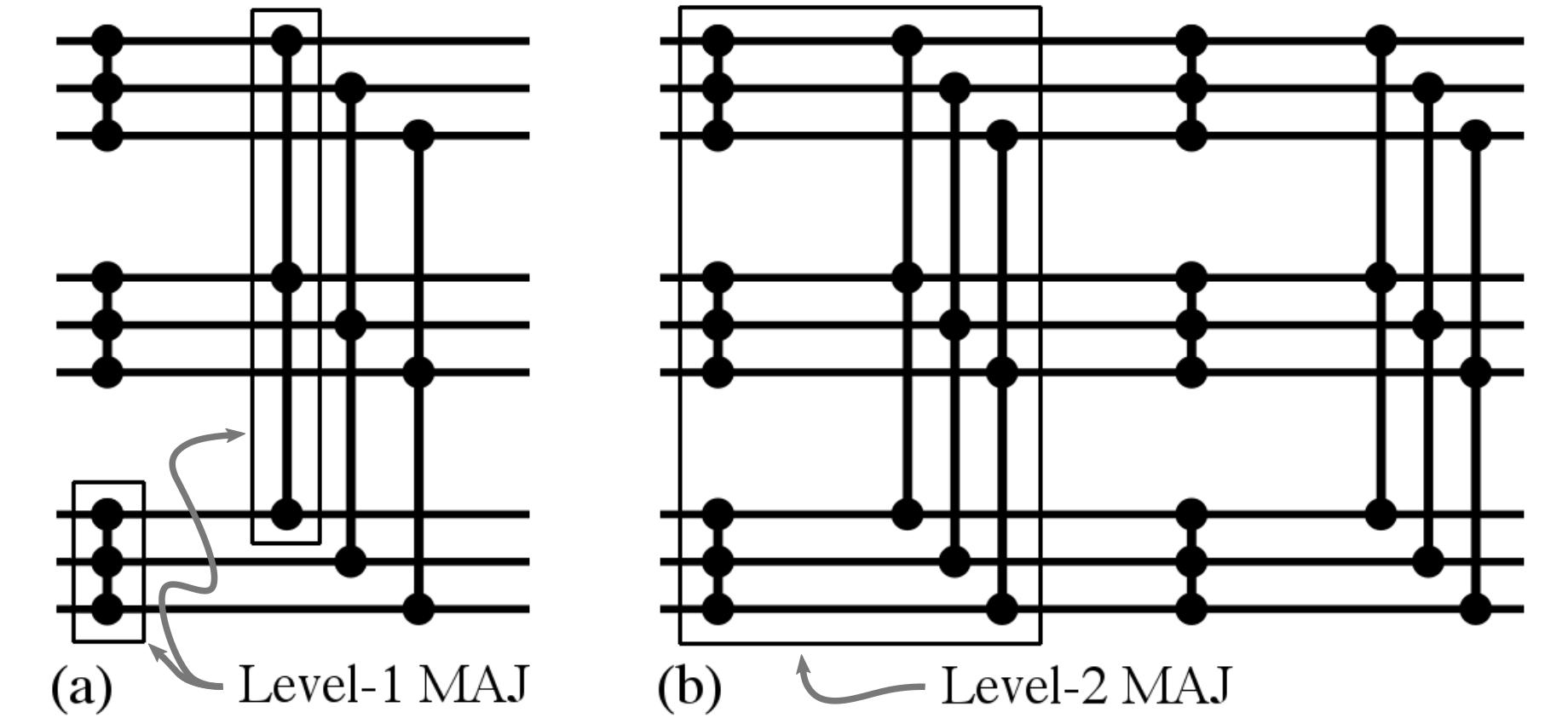}
\caption{Here we show the definition of a level-2 MAJ gate in terms of six level-1 MAJ gates. Each level-1 MAJ gate is depicted as three dots connected by a vertical line. Note that each of the three input lines to a level-1 MAJ gate represents a set of three qubits.} 
\label{figS3} 
\end{figure}

Provided we cannot reduce the basic error rate $\varepsilon$, the only way to obtain a logical decay rate that is significantly lower than $3\varepsilon^2$ is to have more than nine bits in our repetition code. We now consider a code with 27 bits and run three level-1 MAJ gates in parallel on these bits. We then apply another set of three level-1 MAJ gates to correct errors transversally across the output blocks. This set-up is shown explicitly in Fig.~\ref{figS3}a. Note that this wiring configuration is similar to the alternating row-to-column construction that we use to create the level-1 MAJ gate (depicted in Fig.~\ref{figS1}), but this time each input line depicts a ``bundle" of three input bits. 

Intuitively, we would expect to see significant improvement in the logical error rate of the network, as a total failure in any one of the initial level-1 MAJ gates will be corrected by the network on the next step. We will call this network a level-2 MAJ gate. This gate is the circuit for correcting our code with size $n=2$, or $9$ bundles of 3 bits. Level-2 MAJ gates can be linked in series as shown in Fig.~\ref{figS3}b.  

At this point it is useful to give precise definitions for four terms:

\vspace{0.5ex}


A \textbf{\textit{logical error}} will refer to a state of a portion of a network which will lead to an error in every output of the network when the network performs perfectly at each subsequent restorative phase.

\vspace{0.5ex}

A \textbf{\textit{gate failure}} means that every output from a specified gate is incorrect. 

\vspace{0.5ex}

An \textbf{\textit{incipient error}} is any failure in the fundamental MAJ3 gates, that occurs independently at the $n^{th}$ error-correction step, at a rate less than $\varepsilon$. 

\vspace{0.5ex}

A \textbf{\textit{propagated error}} is an error present in the network at the $n^{th}$ error-correction step that was generated by some combination of errors at any \textit{previous} error-correction step. 

\vspace{0.5ex}

We now note that a logical error occurs in the error-correction network of Fig.~\ref{figS3} whenever two level-1 MAJ gates fail simultaneously.  First we analyze how this condition can be created solely with incipient errors. Assuming no propagated errors, three or fewer incipient errors can lead to at most one level-1 MAJ failure. So, we must have at least 4 incipient errors to create a logical error. Of the 126 ways to choose 4 errors in nine MAJ 3 gates, 27 combinations lead to two level-1 MAJ failures. So the probability of a logical error due to simultaneous incipient errors is $<27\varepsilon^4$.  This is $3(3\varepsilon^2)^2$, and exactly the sort of improvement over the level-1 MAJ that one would find with a very high threshold concatenation scheme (in that case the first approximation to the threshold would be one-third, and the actual threshold would be one-half). 

Unfortunately, incipient errors are not the only type of error present in the network, and we must account for the effects of propagated errors as well. Simultaneous groups of fewer than 4 errors can have either no effect on the subsequent gates, or they can lead to a failure of a single level-1 MAJ at the next error-correction step. If there is a failure in a single level-1 MAJ at some error-correction step, due to errors in the previous step, then this changes the probability of a logical error, at that step: only two more incipient errors are required at that error-correction step for a logical error. Furthermore, only a single incipient error at this step is required to generate a single level-1 MAJ failure at the next step. 

The above analysis suggests a way to model the steady state behavior of a series of level-2 MAJ gates. We can consider the error-correction process as having three discrete states, which we will label as A, B, and C, corresponding to the possible propagated errors present at any given step. State A corresponds to no propagated errors, state B to one level-1 MAJ failure from an error in previous steps, and C to a logical error due to errors in previous steps. The error-correction process is a stochastic jump process in which there are constant probabilities for transitions to occur between these states at each step. In Table~\ref{tab1} we give the transition probabilities between states A and B and from each of these to state C. In the expressions in Table~\ref{tab1} we have defined $\gamma \equiv 3\epsilon^2-2\epsilon^3$, which is the probability of at least two incipient errors occurring in the input to a level-1 MAJ. 

\begin{table}[t]
\caption{}
\begin{center}
\begin{tabular}{c|c}
Conditional probability & value \\
\hline
$P(A|A)$  & $1-P(B|A)-P(C|A)$ \\
\hline 
$P(B|A)$ & $3\gamma(1-\gamma)^2$ \\
\hline
$P(C|A)$ & $3\gamma^2 - 2\gamma^3$ \\
\hline
$P(A|B)$ & $(1-\epsilon)^6$ \\
\hline
$P(B|B)$ & $6\epsilon(1-\epsilon)^5+3\epsilon^2(1-\epsilon)^4$ \\ 
\hline
$P(C|B)$ & $1-P(A|B)-P(B|B)$ \\ \hline
\end{tabular}
\end{center}
\label{tab1}
\end{table}%

Once a logical error occurs the error-correction process has the same set of transition rates except that the value of the logical bit has been flipped. Since we are only concerned with the probability, $P(C)$, of a logical bit flip at each step, we solve for the steady state probabilities $P(A)_s$ and $P(B)_s$, assuming that no logical error occurs. This involves simply ignoring the $P(C|A)$ and $P(C|B)$ probabilities, normalizing, and solving for the eigenvectors of the $2\times2$ transition matrix.  Then the steady-state probability of a logical bit flip, denoted in the main text by $p_{\ms{ss}}$, is $P(C)_s = P(A)_s P(C|A)+P(B)_s P(C|B)$. The plot in Fig. 2 of the main paper a shows the inverse of the logical error probability as a function of $\epsilon$. For a comparison, we also include a plot of the inverse of $f(\epsilon) = 6^3\epsilon^4$, which would be the approximate logical error probability of a typical concatenated gate with a threshold of $1/6$ at the second level of concatenation \cite{mikeandike}. This is a theoretical construction, as there is no typical concatenation scheme in existence with a threshold of $1/6$. 

\subsubsection{The correction circuit for $n=3$} 
\label{lev3MAJ}
The wiring described above for a series of level-2 MAJ gates can be represented again in 3-by-3 squares as in Fig.~\ref{figS2}.  Now, however, each of the nine squares in the 3-by-3 matrix denotes a bundle of 3 bits that is one of the inputs to the level-1 MAJ, and the error properties of the cycle are as detailed above. 

The level-1 MAJ gate corrects a ``row'' of three bundles, and the level-2 MAJ gate corrects a square of 9 bundles by applying a level-1 MAJ gate first to each of the three columns of the square and then to the three rows. We can now generalize this procedure to a cube (more generally a hypercube of dimension $n$) that has 3 bundles to a side, and to which we apply $3^{n-1}$ level-1 MAJ gates to each dimension in turn. A hypercube with $n$ dimensions is the  repetition code of the main paper with code size $n$. 

The correction circuit for a cube of 27 bundles ($n=3$) is equivalent to running three level-2 MAJ gates in parallel, and then applying level-1 MAJ gates transversally across their outputs. The circuit acts on 81 individual bits, and we will call it a \textit{level-3 MAJ gate}. The error properties of the level-3 MAJ gate can be modeled in the same way as those of the level-2 MAJ gate. That is, as a stochastic process that jumps between a number of discrete states, in which the states denote different configurations of propagated errors, and the transition probabilities are due to possible combinations of incipient errors. The stochastic process describing the level-3 MAJ is more complicated than that for the level-2 MAJ because there are more possibilities for propagated errors from the level-1 MAJ gates errors which are not equivalent to a logical error, and, moreover, the location of propagated errors in the cycle has an effect on the probabilities of future states. By carefully enumerating the possible configurations of errors we find that the level-3 MAJ error-correction network has seven distinct propagated error states that do not lead immediately to a logical error. The transition probabilities between these seven states can be worked out through straightforward counting. We give further details of this calculation in the next section. The explicit matrix of transition probabilities can be found in the MATLAB function \textit{SteadyStateFailErrorProb}.

As must be the case, none of the rows of the 7-by-7 transition matrix sum to one. The sum of the elements of row $m$ is $1-P_m$, where $P_m$ is the probability of a logical error given that the network is in state $m$.  Let us define the 7-element vector $\mathbf{p}_{\ms{fail}} \equiv (P_1,\ldots,P_7)$. We can obtain the steady-state probabilities for the 7 states, given that a logical error has not occurred, by normalizing each of the rows of the transition matrix and raising the resulting matrix to a sufficiently high power. Denoting the vector of these steady-state probabilities by $\mathbf{p}_{\ms{ss}}$, the steady-state probability of a logical error in any given restorative phase is $P^{\ms{fail}}_{\ms{ss}} = \mathbf{p}_{\ms{ss}} \cdot \mathbf{p}_{\ms{fail}}$. Numerical results indicate that the level-3 MAJ begins to have a beneficial effect on the logical error probability at a threshold value of $\epsilon\approx1/6$, as shown in Figure~3a of the main paper. We find that for $\epsilon = 1\%$ the level-3 MAJ gate achieves a logical error rate of $3\times10^{-11}$ per restorative phase. This is, in fact, very close to what would be the logical error rate of a standard concatenation scheme with a threshold of $1/6$ at the third level of concatenation, namely $6^7\epsilon^8$ \cite{mikeandike}. (Again, this is a theoretical construction, there is no actual concatenation scheme with this threshold.) That our values do not quite match the ideal constant of $1/6$ must be due in no small part to the approximation, not employed by von Neumann, that incipient errors never correct propagated output errors from a level-1 MAJ. We compare the logical error probability to that of two hypothetical concatenated codes, this time at level three, in the plot in Fig. 2b of the main paper. The expressions for the error probabilities of the concatenated codes are $6^7\epsilon^8$ and $7^7\epsilon^8$, corresponding to thresholds of $1/6$ and $1/7$, respectively.

The close relationship between the error probability and $6^7\epsilon^8$ indicates, and we expect, that our wiring scheme can be generalized further to higher dimensions of 3-by-3 hypercubes if necessary, and will achieve the same polylogarithmic scaling between overhead and logical error as obtained by standard concatenation schemes. We also believe the threshold should remain close to $1/6$, as increasing the code size only increases the independence of the gates, bringing it closer to von Neumann's idealization. Formalizing and proving these claims may be a question for future research.   

\subsubsection{Distinct error states of the level-3 MAJ} 
\label{App1}
We now detail the enumeration of the seven states that describe the level-3 MAJ error-correction process, and the calculation of the transition probabilities. Each correction step in the level-3 MAJ error-correction cycle consists of nine level-1 MAJ gates applied in parallel. Each of these gates can fail due to the presence of incipient and/or propagated errors in its inputs, creating a propagated error at the next error-correction step. It is the precise configuration of the error-correction wiring that determines how these propagated errors affect the next error-correction step. Consider the $n^{th}$ step, comprising nine parallel level-1 MAJ gates, as applied down the columns of three distinct 3 by 3 squares. Given a failure in any one of these level-1 MAJ's, on the $n^{th}+1$ step, this propagated error will be input into three different gates. Specifically, a failure of the level-1 MAJ applied on the $k^{th}$ column of the $l^{th}$ square leads to a propagated input error in the position $(row\;l, column\;k)$ in each square in the following step.  The level-1 MAJ's at this step are applied along the rows of the squares. Level-1 MAJ's into which a propagated error have been input have a failure probability $P=2\varepsilon-\varepsilon^2$, which is the probability of at least one incipient error occurring in the other two inputs. Gates which have no propagated input errors have a failure probability  $I = 3\varepsilon^2-2\varepsilon^3$, which is the probability of at least two incipient errors occurring in the three inputs. Two propagated errors input into the same gate lead immediately to a failure of that gate. A failure in a level-1 MAJ across the $k^{th}$ row of the $l^{th}$ square leads to a leads to a propagated input error input into the position $(row\;k, column\;l)$ in each square at the $n^{th}+2$ step. Level-1 MAJ's are applied along the columns here, and the cycle repeats. 

The above implies that there are seven distinct propagated error states at any given error-correction step that have distinct failure probabilities and must thus be treated as distinct states of the system. These states can be identified in terms of a single 3-by-3 square to which three level-1 MAJ gates are applied. This is due to the fact that there are actually three such squares, but all of them are identical at any error-correction step (assuming, as above, that there are no errors that correct the state):

The seven propagated error ``states'' are characterized as the following groupings: 
\begin{enumerate}
\item All states with no propagated errors.
\item All states with 1 propagated error.
\item All states with 2 propagated errors that are not in the same row or column as that to which the level-1 MAJ gate is applied. 
\item All states with 3 propagated errors that are not in the same row or column as that to which the level-1 MAJ gate is applied. 
\item All states with 2 or 3 propagated errors in the same  row or column as that to which the level-1 MAJ gate is applied, and no other propagated errors. 
\item All states with 2 or 3 propagated errors in the same  row or column as that to which the level-1 MAJ gate is applied, and 1 other propagated error.
\item All states with 2 or 3 propagated errors in the same  row or column as that to which the level-1 MAJ gate is applied, and 2 other propagated errors in distinct rows (columns). 

\end{enumerate}

Since 2 sets of 2 or more propagated errors along the direction of the application of a level-1 MAJ gate are equivalent to a logical error, the above list enumerates all the distinct error states of the network.  We have calculated the transition probabilities between all the states exactly using the rules detailed above in the MATLAB function \textit{SteadyStateFailErrorProb}.

\subsection{Unitary implementation and universal FTCC}

\subsubsection{A reversible MAJ gate}

We note that JvN's MAJ gate is not reversible, so we must construct a unitary version of it.  The ``3'' in MAJ3 refers to the fact that it has 3 outputs, and serves to distinguish it from the majority counting gate with one (usable) output, which we call the MAJ1. The MAJ1 has a 3-bit unitary construction. We also introduce another 3-bit unitary gate in which a single bit applies a CNOT operation to two target bits. We will denote this gate by AMP. (This name is an abbreviation of ``amplification'', motivated by the fact that it produces three copies of an initial classical bit.)  The MAJ3 gate consists of a MAJ1 and an AMP gate, and has the minimum number of ancillary input bits required to reversibly implement a three bit majority count with three equivalent outputs. The constructions of these gates are depicted in Fig. 1 of the main paper. 

Since the MAJ3 gate is constructed of more than one elementary gate in this model, we need to determine the error probability $\varepsilon$ for the MAJ3 in terms of a basic gate error probability $p$. Note that the MAJ3 gate will not produce perfectly correlated output bits as JvN assumed for the MAJ gate. However, for the purposes of fault-tolerance perfect correlation is the worst case scenario: a reduction in correlation does not reduce the threshold of JvN's scheme. Thus our MAJ3 gate will act as a MAJ gate with error $\varepsilon$ so long as the error in each output bit is no more than $\varepsilon$. We discuss our error model in the next section, and calculate the error rate of the MAJ3 gate. 

%

\subsubsection{Quantum error model}
\label{SecErrMod}

Error modeling is central to the analysis of quantum computation schemes, but at the same time, it is far from standardized across the field. Varying assumptions about the nature and relative strengths of errors in the circuitry can lead to very different results for an overall estimation of the noise threshold for a scheme. See~\cite{Stephens14} for a discussion off the effects of error assumptions as related to threshold calculations for surface codes. What is standard is that all independent sources of error must be identified, and the effects of these errors be accounted for throughout the relevant section of the circuitry. 

A measurement is always an independent source of error in any quantum computation scheme. Our measurement replacement is a network of preparation locations, three-bit gates, and wires. To work out an estimation of the error rate in a measurement device of this type, we need to assign a failure rate to these three basic processes, and then calculate the failure rate of the larger measurement structure.

Since the outputs of two-bit quantum gates are necessarily correlated, it is common practice to assign two-bit gates a fundamental rate, $p$, for an arbitrary quantum error. This error is then divided up evenly among the possible manifestations of this error in terms of Pauli operators on the individual output bits~\cite{Knill05, Fowler12}. For example the correct output of a CNOT gate would be modified by the operators $IX$, $XX$, or $XY$, etc., each with probability $(1/15)p$, as there are 15 such combinations of Pauli operators. 


Extending this method to the three-bit gates, we see that there are 63 possible Pauli error combinations on three outputs. Since we are only concerned with classical computation in the Z basis, we can ignore the error combinations that contain only Z operators. This leaves 56 error operators, which is $(8/9)p\equiv p_c$, which gives a minimal value for the improvement of a three output quantum gate when used for classical computation. Further improvements may be made by modifying the technology specifically for classical computation. The important definition is that in what follows below, $p$ is defined as the rate of any arbitrary Pauli error combination in the output of a three qubit gate.

The other independent sources of error in the encoding/classical computation procedure are the preparation of $|0\rangle$ and $|1\rangle$ states, and the wires which must transport qubits between gates. We can reasonably assume that these processes are less noisy than a three bit gate, simply because there are fewer qubits involved. The exact relationship in reliability with a three bit gate is of course a function of the actual physical implementation. In this paper we assign a value of $(2/3)p$ to the noise level in the preparation procedures and the wires, based only on the fact that Z ``errors" don't affect this state. It is important to note here, however, that once this assumption has been made, it becomes a requirement of our physical implementation in order to achieve the overall error results presented later. If, however, the preparation and/or wires can be achieved considerably less noisily, then our current results may considerably underestimate the noise tolerance of the overall procedures. 

\subsubsection{Error properties of the gates} 
 
\textit{Errors for} AMP and MAJ1: With the total probability of an intrinsic classical error in a three bit  gate set to $p_c$, the intrinsic probability that exactly one of the three outputs is wrong is $(3/7)p_c$, that exactly two outputs are wrong is $(3/7)p_c$, and that all three outputs are wrong is $(1/7)p_c$. It follows that the total intrinsic probability that a given output bit has an error is $(4/7)p_c$.

\textit{Errors for} MAJ3: The three outputs for the MAJ3 gate are produced by an AMP gate which takes the output of the MAJ1as a control and two $|0\rangle$ states as the targets. An error in each of the input zero states causes a single error in one of the output bits. The difference between these errors and those intrinsic to the AMP gate is that they are independent for the output bits. To simplify calculations in what follows we will over-estimate the errors in the AMP gate by also assigning an independent error to the third output bit that also has probability $(2/3)p$. This simplifies calculations because it makes the errors symmetric in the output bits. Since the MAJ3 gates will be connected to each other in a network with some spacial separation we add an additional factor of $(2/3)p$ to every output of the MAJ3 to account for noise in the wires. The errors in the three outputs of MAJ3 are then obtained by adding the error of MAJ1 to those for the outputs of the final AMP. Thus a given output from the MAJ3 gate will be wrong with probability 

\begin{equation}
   \varepsilon \approx  \left(\frac{2}{3} \right)p + \left(\frac{2}{3} \right)p +\left(\frac{4}{7} + \frac{4}{7}\right) p_c \approx 2.3p . 
\end{equation}

As mentioned above, this significantly overestimates the effect of the error rate $p$ on the logical state of the network. This is due in part to the approximation of simply summing the above probabilities, but also because $\varepsilon$ as defined in Section \ref{secFTCC}  is the probability that \textit{all three} output bits are in error, while here it is the error probability of a \textit{given} output bit. 

\subsubsection{Universal FTCC with reversible gates}
\label{secUniversalComputation}

We now show how classical information can be processed reliably with a threshold of $p=5.5\%$ using reversible gates. We begin by noting that if we input $XY0$ into a MAJ gate the output is equal to $X$ AND $Y$. The AMP gate with a 1 in the control acts as a NOT, and completes a universal set. (If readily available, a basic NOT gate would of course be simpler. Also see~\cite{vonNeumann56} for a clever way to do universal computation with MAJ gates without an explicit NOT gate). The more complex of the two processes is the AND gate, so we can obtain a threshold for fault-tolerant classical computing by examining the action of a logical MAJ gate acting on three coded inputs, $X$,$Y$, and $0$, followed by restorative phases. The threshold is the highest quantum error probability for which a logical MAJ operation, when followed by error correction, can lead to a stable error probability for the code bits. 
 
There are two main differences in the properties of the MAJ gate when it is used for universal computation as opposed to error correction.  When used for error correction, the inputs can be assumed to be the same unless there is an error, which is clearly not the case for computation. For error correcting/restorative purposes, a MAJ gate must have a 1:1 ratio of input to output bits. For universal computation, the ratio is 3:1.  

The logical, computational MAJ gate will be a set of basic MAJ gates which act transversally on 3 coded bits and output 1 coded bit. Our logical, \textit{coded} $X$ and $Y$ bits will each comprise an ensemble of $N$ gates in an error correcting network. For restorative phases, we have used the MAJ3 to maintain $3N$ physical (i.e. fundamental, lowest-level) bits per coded bit. To begin a computational phase, we cut off the last AMP gate from the MAJ3 and output only the useful bit from the MAJ1. This gives us $N$ physical bits per coded bit. These physical bits are wired in parallel to the inputs of a third coded bit, called the \textit{target} bit, which consists of $N$ MAJ3 gates inside each of which a $|0\rangle$ state has already been prepared. The outputs of these MAJ3s give $3N$ physical bits which carry the results of the computation up to some error rate. The correlations between the physical bits will be the same as the correlations previously detailed, and restorative phases can then be implemented as described above. 

The threshold value for the steady state physical bit error rate in an error-correcting network, $\eta$, is $1/2$, which corresponds to a threshold value for $\epsilon$ of $1/6$ in Eq.(\ref{pss1}). This makes sense, as the threshold for any repetition code is $1/2$. To find the threshold for universal computation, we have to look at the basic error rate which will lead to our MAJ3's in the target coded bit having an error rate of $1/2$. 

On the first step of computation, since we don't subject the bits to the final amp gate in the MAJ3, the error rate here is lower than $\epsilon$. We can therefore define a smaller error rate 
\begin{equation}
\epsilon' \equiv \epsilon - (2/3)p-(4/7)p_c =\epsilon/2 
\end{equation}
The other source of error in the bits which are input into the target MAJ3's are the propagated errors in the coded bits. To get the physical bit-wise error probability here, we weight the probabilities of the seven error states detailed in Section~\ref{App1} with the probabilities of selecting an erroneous bit given that state. We call this quantity $\eta$. For details on this calculation, see the MATLAB script \textit{LineErrorScript}, where this variable is called ``propagatedErrorProb." The total error rate in the physical bits which are output from the coded bits and input into the target bit can then be given by
\begin{equation}
p_{in}= \epsilon' + (1-\epsilon')(\eta)
\end{equation}
The probability that a given output bit from a MAJ3 in the target network is incorrect is the probability that either one of the input bits is in error, or the preparation of the $|0\rangle$ state is in error, or this final MAJ3 has an error. This probability, $p_{target}$, can be given by 
\begin{equation}
  p_{target} = (1-(1-p_{in})^2(1-\epsilon)(1-(2/3)p))
\end{equation}
The threshold for universal computation must be the value for $p$ which gives $p_{target} < 1/2$. This in turn gives $p < 5.5\%$ or $\epsilon < 12.9\%$.  

\subsection{Spatial implementation/wire length}

\begin{figure}[!]
\leavevmode\includegraphics[width=0.2\hsize]{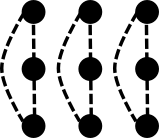}
\caption{Wiring for three parallel level-1 MAJ gates- one third of the Level 3 MAJ.  Dots are MAJ3's. Dashed lines indicate a a two-way signal between gates. Each gate has one internal wire which reroutes a gate output to the input.} 
\label{figW1} 
\end{figure}

\begin{figure}[!]
\leavevmode\includegraphics[width=0.4\hsize]{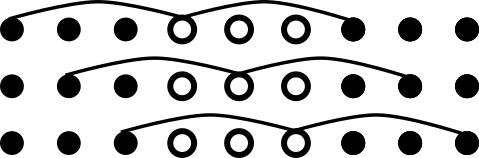}
\caption{One of 3 layers of a computational gate at Level 3. Each cube outputs one wire per gate (MAJ1 output), and the inner (target) cube processes with MAJ3's.} 
\label{figW2} 
\end{figure}

As the level-3 MAJ gate comprises nine level-1 MAJ gates in parallel, it can be implemented on a 3x3x3 array of MAJ3 gates. The wiring in Fig.~\ref{figW1} represents the interaction necessary for any two dimensional slice of the cube at a given time step. Each MAJ3 has an internal output which is rewired as an input and also exchanges outputs with two other gates along the current dimension. This is a restorative phase, and it is repeated as desired sequentially across each dimension of the cube. Universal computation requires an interaction between three such cubes as in Fig.~\ref{figW2}. These figures show that at level three, our wires need not extend further than the length of three spaces in our gate array.  If we want to go further than level 3, we now have to interact three such cubes for restorative purposes, and then interact three of these blocks of eighty-one gates for universal computation. Doing so requires wire lengths of $3\sqrt{2}$ lattice spaces, as shown in Fig. 3b of the main paper. This would necessarily increase the fundamental physical error rate in the network. However, extrapolating from the relationships in Section~\ref{secRestorativePhase}, our fourth-level network should have a logical error rate $\approx6^{15}\epsilon^{16}$, which for $\epsilon=10\%$ should lead to network error $\sim10^{-15}$ per restorative phase. 

\subsection{Replacing measurements with unitary circuits} 

\subsubsection{Encoding with unitary gates}
\label{secEncoding}

The above analysis provides a scheme to perform reliable universal classical computation with quantum gates on logical bits. To use this procedure to process the classical information stored in a single qubit, and thus to emulate a classical measurement and subsequent classical processing, we must encode the single bit into a repetition code. That is, we must make many copies of its basis state.  Naturally we need this encoding to be as noise free as possible. To be more precise, the output from the encoding process must be a state which can be handled by the subsequent error correcting network. 

\begin{figure}[b]
\leavevmode\includegraphics[width=1\hsize]{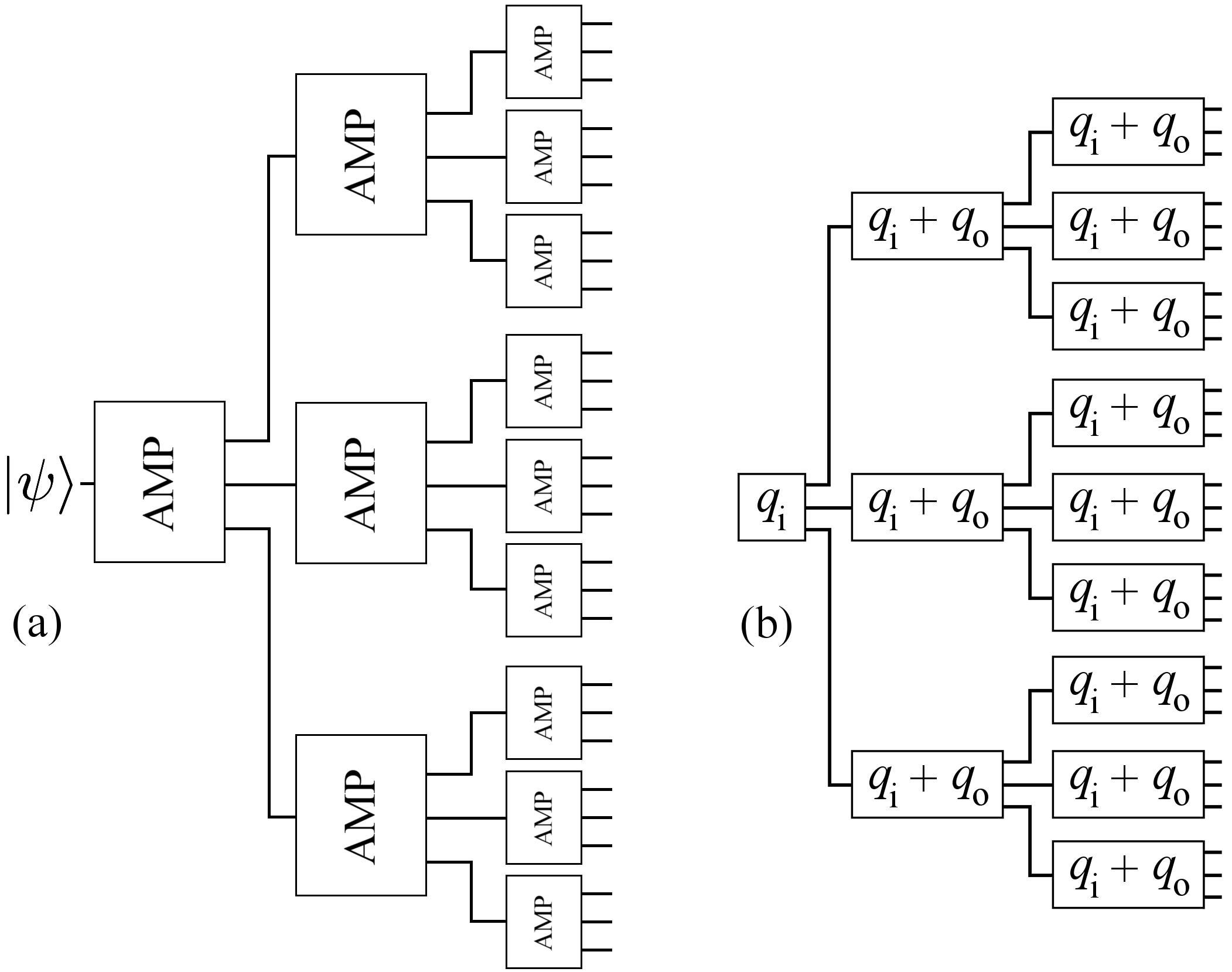}
\caption{(a) Here we show the cascade configuration of AMP gates use to encode a single bit (stored in a single qubit) into a repetition code using $3^{n}$ qubits, where $n\geq 2$ is an integer. (b) The cascade of independent error probabilities that are used to overestimate the errors in the encoding circuit.} 
\label{fig_amp1} 
\end{figure}

Our encoding network is a series of cascaded AMP gates, each of which take as their control input the state we wish to encode, as well as two freshly prepared $|0\rangle$ states as their target inputs (see Fig. 1 of the main paper). The cascade of AMP gates is shown in Fig.\ref{fig_amp1}a.   

The number of output bits blows up exponentially with each time step equal to the execution time of an AMP gate. Clearly, in the absence of noise, our network creates a perfect repetition code of the Z-basis state of our initial bit across the outputs of our AMP gates. With noise, errors will build up during the process, and will be highly correlated across various sections of the code. Therefore, the subsequent error correcting network must be specifically designed to minimize the effects of the correlations. Our error correcting scheme above can be utilized in this way. Our encoding AMP cascade will produce 81 output bits, which can be optimally wired into the error correcting network. Of course, with noisy componentry, there will be some probability that this procedure encodes the wrong state. This is equivalent to both a logical error in the error correcting network and to an erroneous measurement. We find an upper bound on this error probability by counting groups of error locations in our network which lead to a logical error in the subsequent error correcting network. 

The entirety of the encoding procedure is simply to run an AMP cascade for four computational steps. This starts with an interaction with the one bit we wish to ``measure," a location which we call the \textit{initial measurement}. The three outputs of this first AMP are called \textit{level zero} of the encoding process. The nine outputs from the following three AMPs are called \textit{level one}. The twenty-seven outputs from these gates are called \textit{level two}.  At \textit{level three}, we have 81 output bits.  

At this point, we develop an error model in order to find an upper bound on the probability that the wrong state has been encoded. Recall from Section~\ref{SecErrMod} that an AMP gate has the probability $(4/7)p_c$ of producing two or more errors in the output bits. We can pessimistically assume that this is the probability that all three output bits are in error. This has the same effect as an error in the control (code-state) input bit to the AMP gate. So, all bits which are input into the control of an AMP gate are assigned an error probability of $(4/7)p _c \equiv q_{\ms{i}}$. 

Based on the analysis of Section~\ref{SecErrMod}, the probability of a single error in the output of an AMP gate with two $|0\rangle$ states as the target inputs can be overestimated as $4p + (3/7)p_c$. We can further overestimate this error by assigning an \textit{independent} error of $(4/3)p + (1/7)p_c \equiv q_{\ms{o}}$ to each output bit from an AMP gate in our network. All bits which are both inputs and outputs to AMP gates are thus given an error rate of  $q_{\ms{i}}+q_{\ms{o}}= (4/3)p+(5/7)p_c \equiv{ ap}$, where $a \approx 1.97 < 2$. This assignment of independent error probabilities is shown in Fig.\ref{fig_amp1}b.
 
With this model, finding an upper bound on the logical error rate becomes a simple matter of counting the locations in our network which lead to an overall logical error in the code. Under the labels given above, the first order, single-location error rate due to the initial measurement is clearly $q_{\ms{i}}$. At level one, we have three independent error locations with error rate $ap$. The probability that at least two are in error is $3(ap)^2 - 2(ap)^3$. This adds further logical error probability, given no initial measurement error. So, we add the term $(1-q_{\ms{i}})(3(ap)^2 - 2(ap)^3)$. 

\begin{figure}[b]
\leavevmode\includegraphics[width=0.6\hsize]{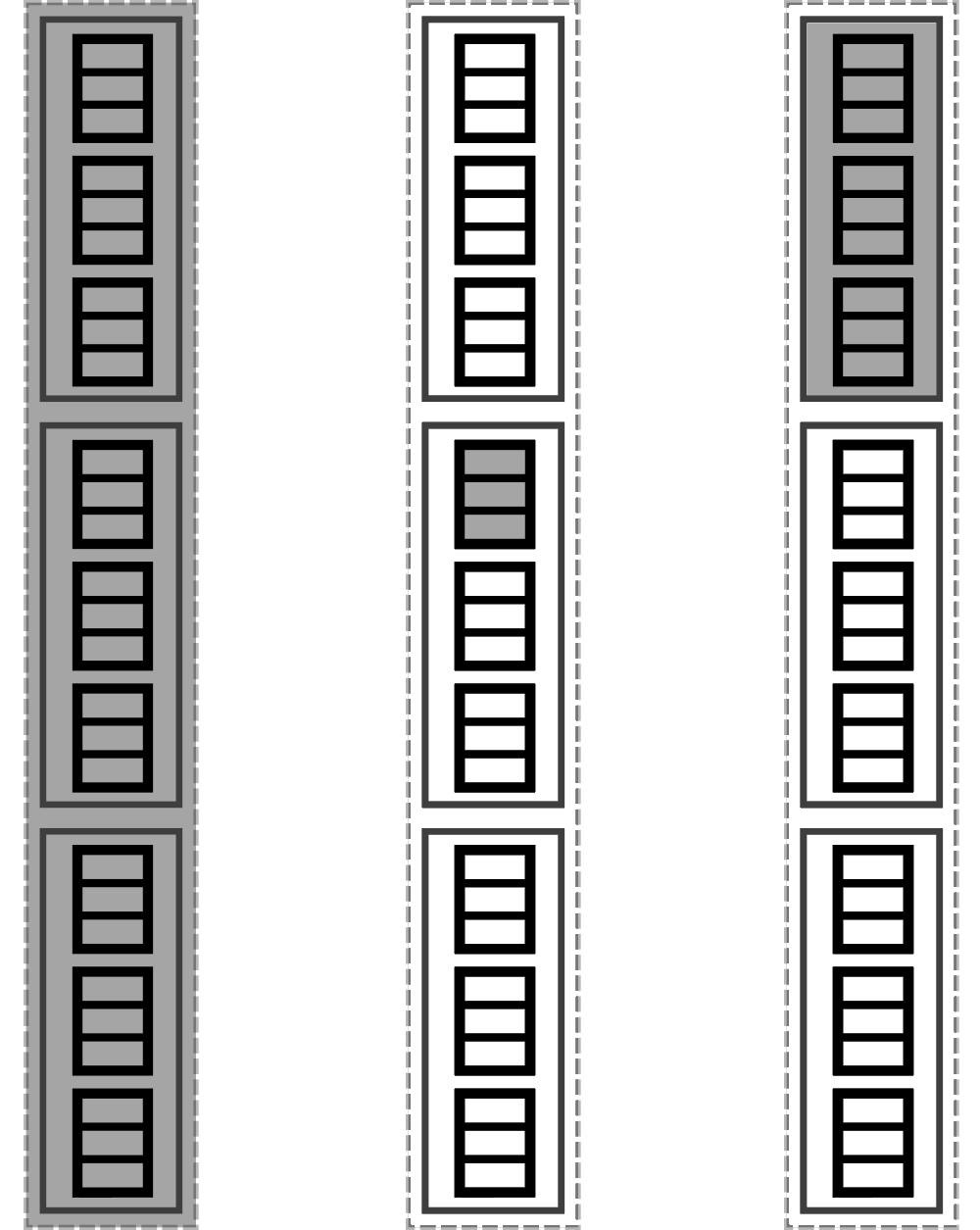}
\caption{A depiction of the groupings of 81 qubit output from the AMP cascade in Fig.~\ref{fig_amp1}. In the first error correction step, 9 level-1 MAJ gates will be applied transversally across the three blocks. The shaded portions indicate blocks of the output which have first-order probability of being in error, due to errors at level zero, two, and one, respectively (from left to right). } 
\label{fig_amp3} 
\end{figure}

Additional sets of error locations will be higher-order, and involve locations at level one. To find this term we have to consider our error correction procedure. When 81 output bits have been created by our cascade, we subject them to 9 transversal level one MAJ gates (or 27 transversal MAJ3s) as detailed above in Section~\ref{lev3MAJ}. The orientation will be such that the each input of the MAJ gates will come from a distinct third of the code. Each distinct third is one of three sections of output bits which descend from the three distinct outputs at level zero. This orientation is shown in Fig.\ref{fig_amp3}.  

With this configuration, we can make the identification that errors at level one in the encoding process correspond to propagated errors in the error correcting network as defined above. Moreover, the combination of errors from levels 2 and 3 are equivalent to incipient errors. Thus, combinations of errors from levels zero and one in the encoding cascade which are not logical errors map directly to the seven error states defined by the error correction circuit (see Section~\ref{App1}), and errors from levels 2 and 3 can be thought of as analogous to $\epsilon$, but at a slightly different rate. The final 27 AMP gates each are assigned the error rate $ap$ as described above. The independent errors in the output bits add a further error probability due to the $|0\rangle$ prep and wire locations. In summation, these errors can be analyzed as incipient errors in the network at the ``zeroth" error correction step at a rate $\epsilon_{0}\approx3.3p=1.4\epsilon$.  


It remains to find the logical error probability due to error sets including level one. Given one error at level zero, the probability of a logical error has the same form as $P(C|B)$ given above.  Namely, $P(L|1_0) = 1-(1-ap)^6-6ap(1-ap)^5-3(ap)^2(1-ap)^4$. Thus the total logical error probability due to error sets with one error at level zero is equal to $(1-q_{\ms{i}})\times(3)(ap)(1-ap)^2\times(P(L|1_0)$.
Finally, given no errors at level zero or one, the logical error rate is $3\alpha^2-2\alpha^3$, where $\alpha = 3(ap)^2-2(ap)^3$, again as above. Combining all terms, the logical error probability due to the AMP cascade can be upper bounded as
\begin{align}
   p_{\ms{fail}} < (4/7)p_c  & + [1-(4/7)p_c]( 3(ap)^2 - 2(ap)^3 \nonumber \\ 
                                                  & + (3)(ap)(1-ap)^2P(L|1_0) \nonumber \\
                                                  & + (1-ap)^3( 3\alpha^2-2\alpha^3) .   \label{pfail} 
\end{align}
For $p = 2.8\%$ this gives $p_{\ms{fail}} = 2.8\%$, and $p_{\ms{fail}} < p$ for all $p<2.8\%$.

The logical error rate for the first few error correction steps will depend on the initial distribution of non-logical error states. This can also be found through straightforward counting procedure which maps error states in the cascade to the seven states of the level-3 MAJ. It is worth mentioning again that the results in this and Section~\ref{secUniversalComputation} are highly dependent on the chosen error model and chosen unitary construction of the 5-bit MAJ3 gate.

\subsection{Decoding the logical state} 
\label{Dec}

Decoding the logical state is the process of placing this state in a single bit. It is important, however, to distinguish between the tasks of placing the logical bit in a noisy mesoscopic bit (something we discuss how to do in the main text), and placing it in a noise-free macroscopic bit, which may be referred to as ``reading out'' the logical bit. For the former, since the final bit cannot avoid having an error of order $p$, which is much larger than the logical error, the process itself introduces an error much larger than the logical error. On the other hand, when we want to place the logical state in a single error-free bit we want to use a procedure that introduces as little additional error as possible. 

The straightforward method would be to measure each qubit, at which point classical error-free majority counting can be used to decode. It remains to ask what happens if the measurements are very noisy. In this case one would make multiple copies of the code bit prior to measuring it, so that there are many physical bits to measure. With enough bits to measure, so long as the measurement error (including the intrinsic bit error) is less than 50\% for each bit, the correct state will be obtained. Copying a logical bit is a simple computation and therefore, if we have achieved a low enough error rate to be sufficient for universal computation, this copying process will not add any significant error probability. The essential point is that the copying operation can be performed using logical gates, which are transversal and have low error rates precisely because of the error correction: copying is achieved with a transversal CNOT, followed by error correction, and this can be repeated as many times as is necessary to obtain the required number of copies.

\subsection{Replacing post-selection with fixed circuits} 
\label{App2} 

A post-selection process is one in which a state of a quantum system (or systems) is prepared in some way, and then measured. Based on the measurement the system is then either used as part of the physical process (e.g. quantum computation) or is discarded and the state prepared again. This process of re-preparation is repeated until the state can be used for the process. Thus, if the process needs multiple such states, one imagines that the post-selection process is running continually and spitting out states that can be used whenever one is generated.

This procedure fits into our construction of a unitary replacement for measurement and feedback as described in the main paper in the section \textit{Replacing Measurements with Unitary Circuits}. To use a quantum state for further processing dependent on a certain a set of measurement results, one needs to apply a set of transversal controlled-swap gates on the state. The control bit will have the value of 1 if the measurement results, which we have been fault-tolerantly processed, indicate the state is suitable.


\begin{thebibliography}{63}%
\makeatletter
\providecommand \@ifxundefined [1]{%
 \@ifx{#1\undefined}
}%
\providecommand \@ifnum [1]{%
 \ifnum #1\expandafter \@firstoftwo
 \else \expandafter \@secondoftwo
 \fi
}%
\providecommand \@ifx [1]{%
 \ifx #1\expandafter \@firstoftwo
 \else \expandafter \@secondoftwo
 \fi
}%
\providecommand \natexlab [1]{#1}%
\providecommand \enquote  [1]{``#1''}%
\providecommand \bibnamefont  [1]{#1}%
\providecommand \bibfnamefont [1]{#1}%
\providecommand \citenamefont [1]{#1}%
\providecommand \href@noop [0]{\@secondoftwo}%
\providecommand \href [0]{\begingroup \@sanitize@url \@href}%
\providecommand \@href[1]{\@@startlink{#1}\@@href}%
\providecommand \@@href[1]{\endgroup#1\@@endlink}%
\providecommand \@sanitize@url [0]{\catcode `\\12\catcode `\$12\catcode
  `\&12\catcode `\#12\catcode `\^12\catcode `\_12\catcode `\%12\relax}%
\providecommand \@@startlink[1]{}%
\providecommand \@@endlink[0]{}%
\providecommand \url  [0]{\begingroup\@sanitize@url \@url }%
\providecommand \@url [1]{\endgroup\@href {#1}{\urlprefix }}%
\providecommand \urlprefix  [0]{URL }%
\providecommand \Eprint [0]{\href }%
\providecommand \doibase [0]{http://dx.doi.org/}%
\providecommand \selectlanguage [0]{\@gobble}%
\providecommand \bibinfo  [0]{\@secondoftwo}%
\providecommand \bibfield  [0]{\@secondoftwo}%
\providecommand \translation [1]{[#1]}%
\providecommand \BibitemOpen [0]{}%
\providecommand \bibitemStop [0]{}%
\providecommand \bibitemNoStop [0]{.\EOS\space}%
\providecommand \EOS [0]{\spacefactor3000\relax}%
\providecommand \BibitemShut  [1]{\csname bibitem#1\endcsname}%
\let\auto@bib@innerbib\@empty
\bibitem [{\citenamefont {von Neumann}(1956)}]{vonNeumann56}%
  \BibitemOpen
  \bibfield  {author} {\bibinfo {author} {\bibfnamefont {J.}~\bibnamefont {von
  Neumann}},\ }in\ \href@noop {} {\emph {\bibinfo {booktitle} {Automata
  studies}}},\ \bibinfo {editor} {edited by\ \bibinfo {editor} {\bibfnamefont
  {C.~E.}\ \bibnamefont {Shannon}}\ and\ \bibinfo {editor} {\bibfnamefont
  {J.}~\bibnamefont {McCarthy}}}\ (\bibinfo  {publisher} {Princeton University
  Press},\ \bibinfo {address} {Princeton, NJ},\ \bibinfo {year} {1956})\ pp.\
  \bibinfo {pages} {43--98}\BibitemShut {NoStop}%
\bibitem [{\citenamefont {Pippenger}(1990)}]{Pippenger90}%
  \BibitemOpen
  \bibfield  {author} {\bibinfo {author} {\bibfnamefont {N.}~\bibnamefont
  {Pippenger}},\ }in\ \href@noop {} {\emph {\bibinfo {booktitle} {Proceedings
  of Symposia in Pure Mathematics, vol. 50: The Legacy of John von Neumann}}},\
  \bibinfo {editor} {edited by\ \bibinfo {editor} {\bibfnamefont
  {J.}~\bibnamefont {Glimm}}, \bibinfo {editor} {\bibfnamefont
  {J.}~\bibnamefont {Impagliazzo}}, \ and\ \bibinfo {editor} {\bibfnamefont
  {I.}~\bibnamefont {Singer}}}\ (\bibinfo  {publisher} {American Mathematical
  Society},\ \bibinfo {address} {Providence, RI},\ \bibinfo {year} {1990})\ p.\
  \bibinfo {pages} {311}\BibitemShut {NoStop}%
\bibitem [{\citenamefont {Nikolic}\ \emph {et~al.}(2002)\citenamefont
  {Nikolic}, \citenamefont {Sadek},\ and\ \citenamefont {Forshaw}}]{Nikolic02}%
  \BibitemOpen
  \bibfield  {author} {\bibinfo {author} {\bibfnamefont {K.}~\bibnamefont
  {Nikolic}}, \bibinfo {author} {\bibfnamefont {A.}~\bibnamefont {Sadek}}, \
  and\ \bibinfo {author} {\bibfnamefont {M.}~\bibnamefont {Forshaw}},\
  }\href@noop {} {\bibfield  {journal} {\bibinfo  {journal} {Nanotechnology}\
  }\textbf {\bibinfo {volume} {13}},\ \bibinfo {pages} {357} (\bibinfo {year}
  {2002})}\BibitemShut {NoStop}%
\bibitem [{\citenamefont {Roy}\ and\ \citenamefont {Beiu}(2005)}]{Roy05}%
  \BibitemOpen
  \bibfield  {author} {\bibinfo {author} {\bibfnamefont {S.}~\bibnamefont
  {Roy}}\ and\ \bibinfo {author} {\bibfnamefont {V.}~\bibnamefont {Beiu}},\
  }\href@noop {} {\bibfield  {journal} {\bibinfo  {journal} {IEEE Transactions
  on Nanotechnology}\ }\textbf {\bibinfo {volume} {4}},\ \bibinfo {pages} {441}
  (\bibinfo {year} {2005})}\BibitemShut {NoStop}%
\bibitem [{\citenamefont {Bhaduri}\ \emph {et~al.}(2007)\citenamefont
  {Bhaduri}, \citenamefont {Shukla}, \citenamefont {Graham},\ and\
  \citenamefont {Gokhale}}]{Bhaduri07}%
  \BibitemOpen
  \bibfield  {author} {\bibinfo {author} {\bibfnamefont {D.}~\bibnamefont
  {Bhaduri}}, \bibinfo {author} {\bibfnamefont {S.}~\bibnamefont {Shukla}},
  \bibinfo {author} {\bibfnamefont {P.}~\bibnamefont {Graham}}, \ and\ \bibinfo
  {author} {\bibfnamefont {M.}~\bibnamefont {Gokhale}},\ }\href@noop {}
  {\bibfield  {journal} {\bibinfo  {journal} {Nanotechnology, IEEE Transactions
  on}\ }\textbf {\bibinfo {volume} {6}},\ \bibinfo {pages} {265} (\bibinfo
  {year} {2007})}\BibitemShut {NoStop}%
\bibitem [{\citenamefont {Beiu}\ \emph {et~al.}(2007)\citenamefont {Beiu},
  \citenamefont {Ibrahim},\ and\ \citenamefont {Lazarova-Molnar}}]{Beiu07}%
  \BibitemOpen
  \bibfield  {author} {\bibinfo {author} {\bibfnamefont {V.}~\bibnamefont
  {Beiu}}, \bibinfo {author} {\bibfnamefont {W.}~\bibnamefont {Ibrahim}}, \
  and\ \bibinfo {author} {\bibfnamefont {S.}~\bibnamefont {Lazarova-Molnar}},\
  }in\ \href@noop {} {\emph {\bibinfo {booktitle} {International
  Work-Conference on Artificial Neural Networks}}}\ (\bibinfo {organization}
  {Springer},\ \bibinfo {year} {2007})\ pp.\ \bibinfo {pages}
  {487--496}\BibitemShut {NoStop}%
\bibitem [{\citenamefont {Ma}\ and\ \citenamefont {Lombardi}(2008)}]{Ma08}%
  \BibitemOpen
  \bibfield  {author} {\bibinfo {author} {\bibfnamefont {X.}~\bibnamefont
  {Ma}}\ and\ \bibinfo {author} {\bibfnamefont {F.}~\bibnamefont {Lombardi}},\
  }in\ \href@noop {} {\emph {\bibinfo {booktitle} {2008 IEEE International
  Symposium on Defect and Fault Tolerance of VLSI Systems}}}\ (\bibinfo
  {organization} {IEEE},\ \bibinfo {year} {2008})\ pp.\ \bibinfo {pages}
  {236--244}\BibitemShut {NoStop}%
\bibitem [{\citenamefont {Han}\ \emph {et~al.}(2011)\citenamefont {Han},
  \citenamefont {Boykin}, \citenamefont {Chen}, \citenamefont {Liang},\ and\
  \citenamefont {Fortes}}]{Han11}%
  \BibitemOpen
  \bibfield  {author} {\bibinfo {author} {\bibfnamefont {J.}~\bibnamefont
  {Han}}, \bibinfo {author} {\bibfnamefont {E.~R.}\ \bibnamefont {Boykin}},
  \bibinfo {author} {\bibfnamefont {H.}~\bibnamefont {Chen}}, \bibinfo {author}
  {\bibfnamefont {J.}~\bibnamefont {Liang}}, \ and\ \bibinfo {author}
  {\bibfnamefont {J.~A.}\ \bibnamefont {Fortes}},\ }\href@noop {} {\bibfield
  {journal} {\bibinfo  {journal} {Nanotechnology, IEEE Transactions on}\
  }\textbf {\bibinfo {volume} {10}},\ \bibinfo {pages} {1099} (\bibinfo {year}
  {2011})}\BibitemShut {NoStop}%
\bibitem [{\citenamefont {Sen}\ \emph {et~al.}(2016)\citenamefont {Sen},
  \citenamefont {Sahu}, \citenamefont {Mukherjee}, \citenamefont {Nath},\ and\
  \citenamefont {Sikdar}}]{Sen16}%
  \BibitemOpen
  \bibfield  {author} {\bibinfo {author} {\bibfnamefont {B.}~\bibnamefont
  {Sen}}, \bibinfo {author} {\bibfnamefont {Y.}~\bibnamefont {Sahu}}, \bibinfo
  {author} {\bibfnamefont {R.}~\bibnamefont {Mukherjee}}, \bibinfo {author}
  {\bibfnamefont {R.~K.}\ \bibnamefont {Nath}}, \ and\ \bibinfo {author}
  {\bibfnamefont {B.~K.}\ \bibnamefont {Sikdar}},\ }\href@noop {} {\bibfield
  {journal} {\bibinfo  {journal} {Microelectronics Journal}\ }\textbf {\bibinfo
  {volume} {47}},\ \bibinfo {pages} {7} (\bibinfo {year} {2016})}\BibitemShut
  {NoStop}%
\bibitem [{\citenamefont {Toom}(1980)}]{Toom80}%
  \BibitemOpen
  \bibfield  {author} {\bibinfo {author} {\bibfnamefont {A.~L.}\ \bibnamefont
  {Toom}},\ }\href@noop {} {\bibfield  {journal} {\bibinfo  {journal} {Advances
  in Probability}\ }\textbf {\bibinfo {volume} {6}},\ \bibinfo {pages} {549}
  (\bibinfo {year} {1980})}\BibitemShut {NoStop}%
\bibitem [{\citenamefont {G{\'a}cs}(1983)}]{Gacs83}%
  \BibitemOpen
  \bibfield  {author} {\bibinfo {author} {\bibfnamefont {P.}~\bibnamefont
  {G{\'a}cs}},\ }in\ \href@noop {} {\emph {\bibinfo {booktitle} {Proceedings of
  the fifteenth annual ACM symposium on Theory of computing}}}\ (\bibinfo
  {organization} {ACM},\ \bibinfo {year} {1983})\ pp.\ \bibinfo {pages}
  {32--41}\BibitemShut {NoStop}%
\bibitem [{\citenamefont {G{\'a}cs}\ and\ \citenamefont {Reif}(1985)}]{Gacs85}%
  \BibitemOpen
  \bibfield  {author} {\bibinfo {author} {\bibfnamefont {P.}~\bibnamefont
  {G{\'a}cs}}\ and\ \bibinfo {author} {\bibfnamefont {J.}~\bibnamefont
  {Reif}},\ }in\ \href@noop {} {\emph {\bibinfo {booktitle} {Proceedings of the
  seventeenth annual ACM symposium on Theory of computing}}}\ (\bibinfo
  {organization} {ACM},\ \bibinfo {year} {1985})\ pp.\ \bibinfo {pages}
  {388--395}\BibitemShut {NoStop}%
\bibitem [{\citenamefont {Han}\ and\ \citenamefont {Jonker}(2003)}]{Han03}%
  \BibitemOpen
  \bibfield  {author} {\bibinfo {author} {\bibfnamefont {J.}~\bibnamefont
  {Han}}\ and\ \bibinfo {author} {\bibfnamefont {P.}~\bibnamefont {Jonker}},\
  }\href@noop {} {\bibfield  {journal} {\bibinfo  {journal} {Nanotechnology}\
  }\textbf {\bibinfo {volume} {14}},\ \bibinfo {pages} {224} (\bibinfo {year}
  {2003})}\BibitemShut {NoStop}%
\bibitem [{\citenamefont {Peper}\ \emph {et~al.}(2004)\citenamefont {Peper},
  \citenamefont {Lee}, \citenamefont {Abo}, \citenamefont {Isokawa},
  \citenamefont {Adachi}, \citenamefont {Matsui},\ and\ \citenamefont
  {Mashiko}}]{Peper04}%
  \BibitemOpen
  \bibfield  {author} {\bibinfo {author} {\bibfnamefont {F.}~\bibnamefont
  {Peper}}, \bibinfo {author} {\bibfnamefont {J.}~\bibnamefont {Lee}}, \bibinfo
  {author} {\bibfnamefont {F.}~\bibnamefont {Abo}}, \bibinfo {author}
  {\bibfnamefont {T.}~\bibnamefont {Isokawa}}, \bibinfo {author} {\bibfnamefont
  {S.}~\bibnamefont {Adachi}}, \bibinfo {author} {\bibfnamefont
  {N.}~\bibnamefont {Matsui}}, \ and\ \bibinfo {author} {\bibfnamefont
  {S.}~\bibnamefont {Mashiko}},\ }\href@noop {} {\bibfield  {journal} {\bibinfo
   {journal} {IEEE Transactions on Nanotechnology}\ }\textbf {\bibinfo {volume}
  {3}},\ \bibinfo {pages} {187} (\bibinfo {year} {2004})}\BibitemShut {NoStop}%
\bibitem [{\citenamefont {Di}\ and\ \citenamefont {Lala}(2008)}]{Di08}%
  \BibitemOpen
  \bibfield  {author} {\bibinfo {author} {\bibfnamefont {J.}~\bibnamefont
  {Di}}\ and\ \bibinfo {author} {\bibfnamefont {P.~K.}\ \bibnamefont {Lala}},\
  }in\ \href@noop {} {\emph {\bibinfo {booktitle} {Emerging
  Nanotechnologies}}}\ (\bibinfo  {publisher} {Springer},\ \bibinfo {year}
  {2008})\ pp.\ \bibinfo {pages} {203--226}\BibitemShut {NoStop}%
\bibitem [{\citenamefont {{\v{Z}}aloudek}\ and\ \citenamefont
  {Sekanina}(2011)}]{Zloudek11}%
  \BibitemOpen
  \bibfield  {author} {\bibinfo {author} {\bibfnamefont {L.}~\bibnamefont
  {{\v{Z}}aloudek}}\ and\ \bibinfo {author} {\bibfnamefont {L.}~\bibnamefont
  {Sekanina}},\ }in\ \href@noop {} {\emph {\bibinfo {booktitle} {International
  Conference on Unconventional Computation}}}\ (\bibinfo {organization}
  {Springer},\ \bibinfo {year} {2011})\ pp.\ \bibinfo {pages}
  {234--245}\BibitemShut {NoStop}%
\bibitem [{\citenamefont {de~Maere}\ and\ \citenamefont
  {Ponselet}(2012)}]{deMaere12}%
  \BibitemOpen
  \bibfield  {author} {\bibinfo {author} {\bibfnamefont {A.}~\bibnamefont
  {de~Maere}}\ and\ \bibinfo {author} {\bibfnamefont {L.}~\bibnamefont
  {Ponselet}},\ }\href@noop {} {\bibfield  {journal} {\bibinfo  {journal} {J.
  Stat. Phys.}\ }\textbf {\bibinfo {volume} {147}},\ \bibinfo {pages} {634}
  (\bibinfo {year} {2012})}\BibitemShut {NoStop}%
\bibitem [{\citenamefont {Ponselet}\ \emph {et~al.}(2013)\citenamefont
  {Ponselet} \emph {et~al.}}]{Ponselet13}%
  \BibitemOpen
  \bibfield  {author} {\bibinfo {author} {\bibfnamefont {L.}~\bibnamefont
  {Ponselet}} \emph {et~al.},\ }\emph {\bibinfo {title} {Phase transitions in
  probabilistic cellular automata}},\ \href@noop {} {Ph.D. thesis},\ \bibinfo
  {school} {UCL} (\bibinfo {year} {2013})\BibitemShut {NoStop}%
\bibitem [{\citenamefont {Mairesse}\ and\ \citenamefont
  {Marcovici}(2014)}]{Mairesse14}%
  \BibitemOpen
  \bibfield  {author} {\bibinfo {author} {\bibfnamefont {J.}~\bibnamefont
  {Mairesse}}\ and\ \bibinfo {author} {\bibfnamefont {I.}~\bibnamefont
  {Marcovici}},\ }\href@noop {} {\bibfield  {journal} {\bibinfo  {journal}
  {Theoretical Computer Science}\ }\textbf {\bibinfo {volume} {559}},\ \bibinfo
  {pages} {42} (\bibinfo {year} {2014})}\BibitemShut {NoStop}%
\bibitem [{\citenamefont {S{\l}owi{\'n}ski}\ and\ \citenamefont
  {MacKay}(2015)}]{Slowinski15}%
  \BibitemOpen
  \bibfield  {author} {\bibinfo {author} {\bibfnamefont {P.}~\bibnamefont
  {S{\l}owi{\'n}ski}}\ and\ \bibinfo {author} {\bibfnamefont {R.~S.}\
  \bibnamefont {MacKay}},\ }\href@noop {} {\bibfield  {journal} {\bibinfo
  {journal} {J. Stat. Phys.}\ }\textbf {\bibinfo {volume} {159}},\ \bibinfo
  {pages} {43} (\bibinfo {year} {2015})}\BibitemShut {NoStop}%
\bibitem [{\citenamefont {Lee}\ \emph {et~al.}(2016)\citenamefont {Lee},
  \citenamefont {Peper}, \citenamefont {Leibnitz},\ and\ \citenamefont
  {Gu}}]{Lee16}%
  \BibitemOpen
  \bibfield  {author} {\bibinfo {author} {\bibfnamefont {J.}~\bibnamefont
  {Lee}}, \bibinfo {author} {\bibfnamefont {F.}~\bibnamefont {Peper}}, \bibinfo
  {author} {\bibfnamefont {K.}~\bibnamefont {Leibnitz}}, \ and\ \bibinfo
  {author} {\bibfnamefont {P.}~\bibnamefont {Gu}},\ }\href@noop {} {\bibfield
  {journal} {\bibinfo  {journal} {Information Sciences}\ }\textbf {\bibinfo
  {volume} {352}},\ \bibinfo {pages} {150} (\bibinfo {year}
  {2016})}\BibitemShut {NoStop}%
\bibitem [{\citenamefont {Shor}(1996)}]{Shor96}%
  \BibitemOpen
  \bibfield  {author} {\bibinfo {author} {\bibfnamefont {P.~W.}\ \bibnamefont
  {Shor}},\ }in\ \href@noop {} {\emph {\bibinfo {booktitle} {Proceedings of the
  37th Annual Symposium on Foundations of Computer Science}}},\ \bibinfo
  {series and number} {FOCS '96}\ (\bibinfo  {publisher} {IEEE Computer
  Society},\ \bibinfo {address} {Washington, DC, USA},\ \bibinfo {year}
  {1996})\ p.~\bibinfo {pages} {56}\BibitemShut {NoStop}%
\bibitem [{\citenamefont {Kitaev}(1997)}]{Kitaev97}%
  \BibitemOpen
  \bibfield  {author} {\bibinfo {author} {\bibfnamefont {A.~Y.}\ \bibnamefont
  {Kitaev}},\ }\href@noop {} {\bibfield  {journal} {\bibinfo  {journal}
  {Russian Math. Surveys}\ }\textbf {\bibinfo {volume} {52}},\ \bibinfo {pages}
  {1191} (\bibinfo {year} {1997})}\BibitemShut {NoStop}%
\bibitem [{\citenamefont {Aharonov}\ and\ \citenamefont
  {Ben-Or}(1998)}]{Aharonov98}%
  \BibitemOpen
  \bibfield  {author} {\bibinfo {author} {\bibfnamefont {D.}~\bibnamefont
  {Aharonov}}\ and\ \bibinfo {author} {\bibfnamefont {M.}~\bibnamefont
  {Ben-Or}},\ }in\ \href@noop {} {\emph {\bibinfo {booktitle} {Proceedings of
  the 29th Annual ACM Symposium on the Theory of Computation}}}\ (\bibinfo
  {publisher} {Association for Computing Machinery},\ \bibinfo {address} {New
  York},\ \bibinfo {year} {1998})\ p.\ \bibinfo {pages} {176}\BibitemShut
  {NoStop}%
\bibitem [{\citenamefont {Knill}\ \emph {et~al.}(1998)\citenamefont {Knill},
  \citenamefont {Laflamme},\ and\ \citenamefont {Zurek}}]{Knill98}%
  \BibitemOpen
  \bibfield  {author} {\bibinfo {author} {\bibfnamefont {E.}~\bibnamefont
  {Knill}}, \bibinfo {author} {\bibfnamefont {R.}~\bibnamefont {Laflamme}}, \
  and\ \bibinfo {author} {\bibfnamefont {W.~H.}\ \bibnamefont {Zurek}},\ }\href
  {\doibase 10.1126/science.279.5349.342} {\bibfield  {journal} {\bibinfo
  {journal} {Science}\ }\textbf {\bibinfo {volume} {279}},\ \bibinfo {pages}
  {342} (\bibinfo {year} {1998})}\BibitemShut {NoStop}%
\bibitem [{\citenamefont {Preskill}(1998)}]{Preskill98}%
  \BibitemOpen
  \bibfield  {author} {\bibinfo {author} {\bibfnamefont {J.}~\bibnamefont
  {Preskill}},\ }\href {\doibase 10.1098/rspa.1998.0167} {\bibfield  {journal}
  {\bibinfo  {journal} {Proceedings of the Royal Society of London. Series A:
  Mathematical, Physical and Engineering Sciences}\ }\textbf {\bibinfo {volume}
  {454}},\ \bibinfo {pages} {385} (\bibinfo {year} {1998})}\BibitemShut
  {NoStop}%
\bibitem [{\citenamefont {Gottesman}(1998)}]{Gottesman98}%
  \BibitemOpen
  \bibfield  {author} {\bibinfo {author} {\bibfnamefont {D.}~\bibnamefont
  {Gottesman}},\ }\href {\doibase 10.1103/PhysRevA.57.127} {\bibfield
  {journal} {\bibinfo  {journal} {Phys. Rev. A}\ }\textbf {\bibinfo {volume}
  {57}},\ \bibinfo {pages} {127} (\bibinfo {year} {1998})}\BibitemShut
  {NoStop}%
\bibitem [{\citenamefont {Steane}(1999)}]{Steane99}%
  \BibitemOpen
  \bibfield  {author} {\bibinfo {author} {\bibfnamefont {A.~M.}\ \bibnamefont
  {Steane}},\ }\href@noop {} {\bibfield  {journal} {\bibinfo  {journal}
  {Nature}\ }\textbf {\bibinfo {volume} {399}},\ \bibinfo {pages} {124}
  (\bibinfo {year} {1999})}\BibitemShut {NoStop}%
\bibitem [{\citenamefont {Knill}(2005)}]{Knill05}%
  \BibitemOpen
  \bibfield  {author} {\bibinfo {author} {\bibfnamefont {M.}~\bibnamefont
  {Knill}},\ }\href@noop {} {\bibfield  {journal} {\bibinfo  {journal}
  {Nature}\ }\textbf {\bibinfo {volume} {434}},\ \bibinfo {pages} {265}
  (\bibinfo {year} {2005})}\BibitemShut {NoStop}%
\bibitem [{\citenamefont {Terhal}\ and\ \citenamefont
  {Burkard}(2005)}]{Terhal05}%
  \BibitemOpen
  \bibfield  {author} {\bibinfo {author} {\bibfnamefont {B.~M.}\ \bibnamefont
  {Terhal}}\ and\ \bibinfo {author} {\bibfnamefont {G.}~\bibnamefont
  {Burkard}},\ }\href {\doibase 10.1103/PhysRevA.71.012336} {\bibfield
  {journal} {\bibinfo  {journal} {Phys. Rev. A}\ }\textbf {\bibinfo {volume}
  {71}},\ \bibinfo {pages} {012336} (\bibinfo {year} {2005})}\BibitemShut
  {NoStop}%
\bibitem [{\citenamefont {Nielsen}\ and\ \citenamefont
  {Dawson}(2005)}]{Nielsen05}%
  \BibitemOpen
  \bibfield  {author} {\bibinfo {author} {\bibfnamefont {M.~A.}\ \bibnamefont
  {Nielsen}}\ and\ \bibinfo {author} {\bibfnamefont {C.~M.}\ \bibnamefont
  {Dawson}},\ }\href {\doibase 10.1103/PhysRevA.71.042323} {\bibfield
  {journal} {\bibinfo  {journal} {Phys. Rev. A}\ }\textbf {\bibinfo {volume}
  {71}},\ \bibinfo {pages} {042323} (\bibinfo {year} {2005})}\BibitemShut
  {NoStop}%
\bibitem [{\citenamefont {Aliferis}\ and\ \citenamefont
  {Leung}(2006)}]{Aliferis06}%
  \BibitemOpen
  \bibfield  {author} {\bibinfo {author} {\bibfnamefont {P.}~\bibnamefont
  {Aliferis}}\ and\ \bibinfo {author} {\bibfnamefont {D.~W.}\ \bibnamefont
  {Leung}},\ }\href {\doibase 10.1103/PhysRevA.73.032308} {\bibfield  {journal}
  {\bibinfo  {journal} {Phys. Rev. A}\ }\textbf {\bibinfo {volume} {73}},\
  \bibinfo {pages} {032308} (\bibinfo {year} {2006})}\BibitemShut {NoStop}%
\bibitem [{\citenamefont {Szkopek}\ \emph {et~al.}(2006)\citenamefont
  {Szkopek}, \citenamefont {Boykin}, \citenamefont {Fan}, \citenamefont
  {Roychowdhury}, \citenamefont {Yablonovitch}, \citenamefont {Simms},
  \citenamefont {Gyure},\ and\ \citenamefont {Fong.}}]{Szkopek06}%
  \BibitemOpen
  \bibfield  {author} {\bibinfo {author} {\bibfnamefont {T.}~\bibnamefont
  {Szkopek}}, \bibinfo {author} {\bibfnamefont {P.}~\bibnamefont {Boykin}},
  \bibinfo {author} {\bibfnamefont {H.}~\bibnamefont {Fan}}, \bibinfo {author}
  {\bibfnamefont {V.}~\bibnamefont {Roychowdhury}}, \bibinfo {author}
  {\bibfnamefont {E.}~\bibnamefont {Yablonovitch}}, \bibinfo {author}
  {\bibfnamefont {G.}~\bibnamefont {Simms}}, \bibinfo {author} {\bibfnamefont
  {M.}~\bibnamefont {Gyure}}, \ and\ \bibinfo {author} {\bibfnamefont
  {B.}~\bibnamefont {Fong.}},\ }\href@noop {} {\bibfield  {journal} {\bibinfo
  {journal} {IEEE Trans. Nano.}\ }\textbf {\bibinfo {volume} {5}},\ \bibinfo
  {pages} {42} (\bibinfo {year} {2006})}\BibitemShut {NoStop}%
\bibitem [{\citenamefont {Aliferis}\ \emph {et~al.}(2006)\citenamefont
  {Aliferis}, \citenamefont {Gottesman},\ and\ \citenamefont
  {Preskill}}]{Aliferis06b}%
  \BibitemOpen
  \bibfield  {author} {\bibinfo {author} {\bibfnamefont {P.}~\bibnamefont
  {Aliferis}}, \bibinfo {author} {\bibfnamefont {D.}~\bibnamefont {Gottesman}},
  \ and\ \bibinfo {author} {\bibfnamefont {J.}~\bibnamefont {Preskill}},\
  }\href@noop {} {\bibfield  {journal} {\bibinfo  {journal} {Quant. Inf.
  Comput.}\ }\textbf {\bibinfo {volume} {6}},\ \bibinfo {pages} {97} (\bibinfo
  {year} {2006})}\BibitemShut {NoStop}%
\bibitem [{\citenamefont {Dawson}\ \emph {et~al.}(2006)\citenamefont {Dawson},
  \citenamefont {Haselgrove},\ and\ \citenamefont {Nielsen}}]{Dawson06}%
  \BibitemOpen
  \bibfield  {author} {\bibinfo {author} {\bibfnamefont {C.~M.}\ \bibnamefont
  {Dawson}}, \bibinfo {author} {\bibfnamefont {H.~L.}\ \bibnamefont
  {Haselgrove}}, \ and\ \bibinfo {author} {\bibfnamefont {M.~A.}\ \bibnamefont
  {Nielsen}},\ }\href {\doibase 10.1103/PhysRevLett.96.020501} {\bibfield
  {journal} {\bibinfo  {journal} {Phys. Rev. Lett.}\ }\textbf {\bibinfo
  {volume} {96}},\ \bibinfo {pages} {020501} (\bibinfo {year}
  {2006})}\BibitemShut {NoStop}%
\bibitem [{\citenamefont {Svore}\ \emph {et~al.}(2007)\citenamefont {Svore},
  \citenamefont {Divincenzo},\ and\ \citenamefont {Terhal}}]{Svore07}%
  \BibitemOpen
  \bibfield  {author} {\bibinfo {author} {\bibfnamefont {K.~M.}\ \bibnamefont
  {Svore}}, \bibinfo {author} {\bibfnamefont {D.~P.}\ \bibnamefont
  {Divincenzo}}, \ and\ \bibinfo {author} {\bibfnamefont {B.~M.}\ \bibnamefont
  {Terhal}},\ }\href@noop {} {\bibfield  {journal} {\bibinfo  {journal} {Quant.
  Inf. Comput.}\ }\textbf {\bibinfo {volume} {7}},\ \bibinfo {pages} {297}
  (\bibinfo {year} {2007})}\BibitemShut {NoStop}%
\bibitem [{\citenamefont {Aharonov}\ and\ \citenamefont
  {Ben-Or}(2008)}]{Aharonov08}%
  \BibitemOpen
  \bibfield  {author} {\bibinfo {author} {\bibfnamefont {D.}~\bibnamefont
  {Aharonov}}\ and\ \bibinfo {author} {\bibfnamefont {M.}~\bibnamefont
  {Ben-Or}},\ }\href {\doibase 10.1137/S0097539799359385} {\bibfield  {journal}
  {\bibinfo  {journal} {SIAM Journal on Computing}\ }\textbf {\bibinfo {volume}
  {38}},\ \bibinfo {pages} {1207} (\bibinfo {year} {2008})}\BibitemShut
  {NoStop}%
\bibitem [{\citenamefont {Aliferis}\ \emph {et~al.}(2008)\citenamefont
  {Aliferis}, \citenamefont {Gottesman},\ and\ \citenamefont
  {Preskill}}]{Aliferis08}%
  \BibitemOpen
  \bibfield  {author} {\bibinfo {author} {\bibfnamefont {P.}~\bibnamefont
  {Aliferis}}, \bibinfo {author} {\bibfnamefont {D.}~\bibnamefont {Gottesman}},
  \ and\ \bibinfo {author} {\bibfnamefont {J.}~\bibnamefont {Preskill}},\
  }\href@noop {} {\bibfield  {journal} {\bibinfo  {journal} {Quant. Inf.
  Comput.}\ }\textbf {\bibinfo {volume} {8}},\ \bibinfo {pages} {181} (\bibinfo
  {year} {2008})}\BibitemShut {NoStop}%
\bibitem [{\citenamefont {Fujii}\ and\ \citenamefont
  {Tokunaga}(2010)}]{Fujii10}%
  \BibitemOpen
  \bibfield  {author} {\bibinfo {author} {\bibfnamefont {K.}~\bibnamefont
  {Fujii}}\ and\ \bibinfo {author} {\bibfnamefont {Y.}~\bibnamefont
  {Tokunaga}},\ }\href {\doibase 10.1103/PhysRevLett.105.250503} {\bibfield
  {journal} {\bibinfo  {journal} {Phys. Rev. Lett.}\ }\textbf {\bibinfo
  {volume} {105}},\ \bibinfo {pages} {250503} (\bibinfo {year}
  {2010})}\BibitemShut {NoStop}%
\bibitem [{\citenamefont {Koenig}\ \emph {et~al.}(2010)\citenamefont {Koenig},
  \citenamefont {Kuperberg},\ and\ \citenamefont {Reichardt}}]{Koenig10}%
  \BibitemOpen
  \bibfield  {author} {\bibinfo {author} {\bibfnamefont {R.}~\bibnamefont
  {Koenig}}, \bibinfo {author} {\bibfnamefont {G.}~\bibnamefont {Kuperberg}}, \
  and\ \bibinfo {author} {\bibfnamefont {B.~W.}\ \bibnamefont {Reichardt}},\
  }\href {\doibase http://dx.doi.org/10.1016/j.aop.2010.08.001} {\bibfield
  {journal} {\bibinfo  {journal} {Annals of Physics}\ }\textbf {\bibinfo
  {volume} {325}},\ \bibinfo {pages} {2707 } (\bibinfo {year}
  {2010})}\BibitemShut {NoStop}%
\bibitem [{\citenamefont {Paetznick}\ and\ \citenamefont
  {Reichardt}(2013)}]{Paetznick13}%
  \BibitemOpen
  \bibfield  {author} {\bibinfo {author} {\bibfnamefont {A.}~\bibnamefont
  {Paetznick}}\ and\ \bibinfo {author} {\bibfnamefont {B.~W.}\ \bibnamefont
  {Reichardt}},\ }\href {\doibase 10.1103/PhysRevLett.111.090505} {\bibfield
  {journal} {\bibinfo  {journal} {Phys. Rev. Lett.}\ }\textbf {\bibinfo
  {volume} {111}},\ \bibinfo {pages} {090505} (\bibinfo {year}
  {2013})}\BibitemShut {NoStop}%
\bibitem [{\citenamefont {Jones}(2013)}]{Jones13}%
  \BibitemOpen
  \bibfield  {author} {\bibinfo {author} {\bibfnamefont {C.}~\bibnamefont
  {Jones}},\ }\href {\doibase 10.1103/PhysRevA.87.042305} {\bibfield  {journal}
  {\bibinfo  {journal} {Phys. Rev. A}\ }\textbf {\bibinfo {volume} {87}},\
  \bibinfo {pages} {042305} (\bibinfo {year} {2013})}\BibitemShut {NoStop}%
\bibitem [{\citenamefont {Jochym-O'Connor}\ and\ \citenamefont
  {Laflamme}(2014)}]{Jochym14}%
  \BibitemOpen
  \bibfield  {author} {\bibinfo {author} {\bibfnamefont {T.}~\bibnamefont
  {Jochym-O'Connor}}\ and\ \bibinfo {author} {\bibfnamefont {R.}~\bibnamefont
  {Laflamme}},\ }\href {\doibase 10.1103/PhysRevLett.112.010505} {\bibfield
  {journal} {\bibinfo  {journal} {Phys. Rev. Lett.}\ }\textbf {\bibinfo
  {volume} {112}},\ \bibinfo {pages} {010505} (\bibinfo {year}
  {2014})}\BibitemShut {NoStop}%
\bibitem [{\citenamefont {Stephens}(2014)}]{Stephens14}%
  \BibitemOpen
  \bibfield  {author} {\bibinfo {author} {\bibfnamefont {A.~M.}\ \bibnamefont
  {Stephens}},\ }\href {\doibase 10.1103/PhysRevA.89.022321} {\bibfield
  {journal} {\bibinfo  {journal} {Phys. Rev. A}\ }\textbf {\bibinfo {volume}
  {89}},\ \bibinfo {pages} {022321} (\bibinfo {year} {2014})}\BibitemShut
  {NoStop}%
\bibitem [{\citenamefont {Campbell}(2014)}]{Campbell14}%
  \BibitemOpen
  \bibfield  {author} {\bibinfo {author} {\bibfnamefont {E.~T.}\ \bibnamefont
  {Campbell}},\ }\href {\doibase 10.1103/PhysRevLett.113.230501} {\bibfield
  {journal} {\bibinfo  {journal} {Phys. Rev. Lett.}\ }\textbf {\bibinfo
  {volume} {113}},\ \bibinfo {pages} {230501} (\bibinfo {year}
  {2014})}\BibitemShut {NoStop}%
\bibitem [{\citenamefont {Bomb\'{\i}n}(2015)}]{Bombin15}%
  \BibitemOpen
  \bibfield  {author} {\bibinfo {author} {\bibfnamefont {H.}~\bibnamefont
  {Bomb\'{\i}n}},\ }\href {\doibase 10.1103/PhysRevX.5.031043} {\bibfield
  {journal} {\bibinfo  {journal} {Phys. Rev. X}\ }\textbf {\bibinfo {volume}
  {5}},\ \bibinfo {pages} {031043} (\bibinfo {year} {2015})}\BibitemShut
  {NoStop}%
\bibitem [{\citenamefont {Roychowdhury}\ and\ \citenamefont
  {Boykin}(2005)}]{Boykin05}%
  \BibitemOpen
  \bibfield  {author} {\bibinfo {author} {\bibfnamefont {V.~P.}\ \bibnamefont
  {Roychowdhury}}\ and\ \bibinfo {author} {\bibfnamefont {P.~O.}\ \bibnamefont
  {Boykin}},\ }in\ \href {\doibase 10.1109/DSN.2005.83} {\emph {\bibinfo
  {booktitle} {2005 International Conference on Dependable Systems and Networks
  (DSN'05)}}}\ (\bibinfo  {publisher} {IEEE Computer Society},\ \bibinfo
  {address} {Los Alamitos, CA},\ \bibinfo {year} {2005})\ pp.\ \bibinfo {pages}
  {444--453}\BibitemShut {NoStop}%
\bibitem [{\citenamefont {Antonio}\ \emph {et~al.}(2015)\citenamefont
  {Antonio}, \citenamefont {Randall}, \citenamefont {Hensinger}, \citenamefont
  {Morley},\ and\ \citenamefont {Bose}}]{Antonio16x}%
  \BibitemOpen
  \bibfield  {author} {\bibinfo {author} {\bibfnamefont {B.}~\bibnamefont
  {Antonio}}, \bibinfo {author} {\bibfnamefont {J.}~\bibnamefont {Randall}},
  \bibinfo {author} {\bibfnamefont {W.~K.}\ \bibnamefont {Hensinger}}, \bibinfo
  {author} {\bibfnamefont {G.~W.}\ \bibnamefont {Morley}}, \ and\ \bibinfo
  {author} {\bibfnamefont {S.}~\bibnamefont {Bose}},\ }\href@noop {} {\enquote
  {\bibinfo {title} {Classical computation by quantum bits},}\ }\bibinfo
  {howpublished} {arXiv:1509.03420} (\bibinfo {year} {2015})\BibitemShut
  {NoStop}%
\bibitem [{Note1()}]{Note1}%
  \BibitemOpen
  \bibinfo {note} {Calculations of the error rates for the gates,
  error-correction circuits, and universal computation, as well as a brief
  discussion of postselection, are given in the supplemental material, which begins on page 5 of this document. The supplemental material and the main text share this bibliography }\BibitemShut {NoStop}%
\bibitem [{\citenamefont {Jacobs}(2014)}]{Jacobs14}%
  \BibitemOpen
  \bibfield  {author} {\bibinfo {author} {\bibfnamefont {K.}~\bibnamefont
  {Jacobs}},\ }\href@noop {} {\emph {\bibinfo {title} {Quantum measurement
  theory and its applications}}}\ (\bibinfo  {publisher} {Cambridge University
  Press},\ \bibinfo {address} {Cambridge},\ \bibinfo {year} {2014})\BibitemShut
  {NoStop}%
\bibitem [{\citenamefont {Fowler}\ \emph {et~al.}(2009)\citenamefont {Fowler},
  \citenamefont {Stephens},\ and\ \citenamefont {Groszkowski}}]{Fowler09}%
  \BibitemOpen
  \bibfield  {author} {\bibinfo {author} {\bibfnamefont {A.~G.}\ \bibnamefont
  {Fowler}}, \bibinfo {author} {\bibfnamefont {A.~M.}\ \bibnamefont
  {Stephens}}, \ and\ \bibinfo {author} {\bibfnamefont {P.}~\bibnamefont
  {Groszkowski}},\ }\href {\doibase 10.1103/PhysRevA.80.052312} {\bibfield
  {journal} {\bibinfo  {journal} {Phys. Rev. A}\ }\textbf {\bibinfo {volume}
  {80}},\ \bibinfo {pages} {052312} (\bibinfo {year} {2009})}\BibitemShut
  {NoStop}%
\bibitem [{\citenamefont {DiVincenzo}\ and\ \citenamefont
  {Aliferis}(2007)}]{DiVincenzo07}%
  \BibitemOpen
  \bibfield  {author} {\bibinfo {author} {\bibfnamefont {D.~P.}\ \bibnamefont
  {DiVincenzo}}\ and\ \bibinfo {author} {\bibfnamefont {P.}~\bibnamefont
  {Aliferis}},\ }\href {\doibase 10.1103/PhysRevLett.98.020501} {\bibfield
  {journal} {\bibinfo  {journal} {Phys. Rev. Lett.}\ }\textbf {\bibinfo
  {volume} {98}},\ \bibinfo {pages} {020501} (\bibinfo {year}
  {2007})}\BibitemShut {NoStop}%
\bibitem [{\citenamefont {Paz-Silva}\ \emph {et~al.}(2010)\citenamefont
  {Paz-Silva}, \citenamefont {Brennen},\ and\ \citenamefont {Twamley}}]{Paz10}%
  \BibitemOpen
  \bibfield  {author} {\bibinfo {author} {\bibfnamefont {G.~A.}\ \bibnamefont
  {Paz-Silva}}, \bibinfo {author} {\bibfnamefont {G.~K.}\ \bibnamefont
  {Brennen}}, \ and\ \bibinfo {author} {\bibfnamefont {J.}~\bibnamefont
  {Twamley}},\ }\href {\doibase 10.1103/PhysRevLett.105.100501} {\bibfield
  {journal} {\bibinfo  {journal} {Phys. Rev. Lett.}\ }\textbf {\bibinfo
  {volume} {105}},\ \bibinfo {pages} {100501} (\bibinfo {year}
  {2010})}\BibitemShut {NoStop}%
\bibitem [{\citenamefont {Fitzsimons}\ and\ \citenamefont
  {Twamley}(2009)}]{Fitz09}%
  \BibitemOpen
  \bibfield  {author} {\bibinfo {author} {\bibfnamefont {J.}~\bibnamefont
  {Fitzsimons}}\ and\ \bibinfo {author} {\bibfnamefont {J.}~\bibnamefont
  {Twamley}},\ }\href {\doibase https://doi.org/10.1016/j.entcs.2009.12.012}
  {\bibfield  {journal} {\bibinfo  {journal} {Electronic Notes in Theoretical
  Computer Science}\ }\textbf {\bibinfo {volume} {258}},\ \bibinfo {pages} {35
  } (\bibinfo {year} {2009})},\ \bibinfo {note} {proceedings of the Workshop on
  Logical Aspects of Fault Tolerance (LAFT 2009)}\BibitemShut {NoStop}%
\bibitem [{\citenamefont {Fujii}\ \emph {et~al.}(2014)\citenamefont {Fujii},
  \citenamefont {Negoro}, \citenamefont {Imoto},\ and\ \citenamefont
  {Kitagawa}}]{Fujii14}%
  \BibitemOpen
  \bibfield  {author} {\bibinfo {author} {\bibfnamefont {K.}~\bibnamefont
  {Fujii}}, \bibinfo {author} {\bibfnamefont {M.}~\bibnamefont {Negoro}},
  \bibinfo {author} {\bibfnamefont {N.}~\bibnamefont {Imoto}}, \ and\ \bibinfo
  {author} {\bibfnamefont {M.}~\bibnamefont {Kitagawa}},\ }\href {\doibase
  10.1103/PhysRevX.4.041039} {\bibfield  {journal} {\bibinfo  {journal} {Phys.
  Rev. X}\ }\textbf {\bibinfo {volume} {4}},\ \bibinfo {pages} {041039}
  (\bibinfo {year} {2014})}\BibitemShut {NoStop}%
\bibitem [{\citenamefont {Herold}\ \emph {et~al.}(2015)\citenamefont {Herold},
  \citenamefont {Campbell}, \citenamefont {Eisert},\ and\ \citenamefont
  {Kastoryano}}]{Herold15}%
  \BibitemOpen
  \bibfield  {author} {\bibinfo {author} {\bibfnamefont {M.}~\bibnamefont
  {Herold}}, \bibinfo {author} {\bibfnamefont {E.~T.}\ \bibnamefont
  {Campbell}}, \bibinfo {author} {\bibfnamefont {J.}~\bibnamefont {Eisert}}, \
  and\ \bibinfo {author} {\bibfnamefont {M.~J.}\ \bibnamefont {Kastoryano}},\
  }\href@noop {} {\bibfield  {journal} {\bibinfo  {journal} {Npj Quantum
  Information}\ }\textbf {\bibinfo {volume} {1}},\ \bibinfo {pages} {15010}
  (\bibinfo {year} {2015})}\BibitemShut {NoStop}%
\bibitem [{\citenamefont {Nielsen}\ and\ \citenamefont
  {Chuang}(2000)}]{mikeandike}%
  \BibitemOpen
  \bibfield  {author} {\bibinfo {author} {\bibfnamefont {M.~A.}\ \bibnamefont
  {Nielsen}}\ and\ \bibinfo {author} {\bibfnamefont {I.~L.}\ \bibnamefont
  {Chuang}},\ }\href@noop {} {\emph {\bibinfo {title} {Quantum Computation and
  Quantum Information}}}\ (\bibinfo  {publisher} {Cambridge University Press},\
  \bibinfo {address} {Cambridge},\ \bibinfo {year} {2000})\BibitemShut
  {NoStop}%
\bibitem [{\citenamefont {Fowler}\ \emph {et~al.}(2012)\citenamefont {Fowler},
  \citenamefont {Mariantoni}, \citenamefont {Martinis},\ and\ \citenamefont
  {Cleland}}]{Fowler12}%
  \BibitemOpen
  \bibfield  {author} {\bibinfo {author} {\bibfnamefont {A.~G.}\ \bibnamefont
  {Fowler}}, \bibinfo {author} {\bibfnamefont {M.}~\bibnamefont {Mariantoni}},
  \bibinfo {author} {\bibfnamefont {J.~M.}\ \bibnamefont {Martinis}}, \ and\
  \bibinfo {author} {\bibfnamefont {A.~N.}\ \bibnamefont {Cleland}},\ }\href
  {\doibase 10.1103/PhysRevA.86.032324} {\bibfield  {journal} {\bibinfo
  {journal} {Phys. Rev. A}\ }\textbf {\bibinfo {volume} {86}},\ \bibinfo
  {pages} {032324} (\bibinfo {year} {2012})}\BibitemShut {NoStop}%
\bibitem [{Note2()}]{Note2}%
  \BibitemOpen
  \bibinfo {note} {It should be noted that $\varepsilon $ (along with its
  threshold) is more fundamental to the MAJ3 network than $p$. The former is an
  error rate for any module which acts as a MAJ3. The latter depends on our
  unitary implementation of the MAJ3 and our quantum error model~\cite
  {Note1}}\BibitemShut {NoStop}%
\bibitem [{Note3()}]{Note3}%
  \BibitemOpen
  \bibinfo {note} {Decoding the logical state (reading it out) is not required
  for our application here, but we treat it in the supplement~\cite
  {Note1}.}\BibitemShut {Stop}%
\bibitem [{Note4()}]{Note4}%
  \BibitemOpen
  \bibinfo {note} {For $n=3$ with the qubits arranged in a square, the maximum
  number of qubits that lie between any two that must interact is 5, always
  along straight lines in the array. If three dimensions are utilized, this
  same maximum distance still applies for universal computation at $n=4$, which
  can be accomplished with a $9\times 9\times 9$ array.}\BibitemShut {Stop}%
\bibitem [{\citenamefont {Lee}\ and\ \citenamefont {Peper}(2008)}]{Lee08}%
  \BibitemOpen
  \bibfield  {author} {\bibinfo {author} {\bibfnamefont {J.}~\bibnamefont
  {Lee}}\ and\ \bibinfo {author} {\bibfnamefont {F.}~\bibnamefont {Peper}},\
  }in\ \href@noop {} {\emph {\bibinfo {booktitle} {Automata-2008: Theory and
  Applications of Cellular Automata}}},\ \bibinfo {editor} {edited by\ \bibinfo
  {editor} {\bibfnamefont {A.}~\bibnamefont {Adamatzky}}, \bibinfo {editor}
  {\bibfnamefont {R.}~\bibnamefont {Alonso-Sanz}}, \bibinfo {editor}
  {\bibfnamefont {A.}~\bibnamefont {Lawniczak}}, \bibinfo {editor}
  {\bibfnamefont {J.}~\bibnamefont {Martinez}}, \bibinfo {editor}
  {\bibfnamefont {K.}~\bibnamefont {Morita}}, \ and\ \bibinfo {editor}
  {\bibfnamefont {T.}~\bibnamefont {Worsch}}}\ (\bibinfo  {publisher} {Luniver
  Press},\ \bibinfo {address} {Frome, UK},\ \bibinfo {year} {2008})\ p.\
  \bibinfo {pages} {278}\BibitemShut {NoStop}%
\bibitem [{\citenamefont {Hoe}(2010)}]{Hoe10}%
  \BibitemOpen
  \bibfield  {author} {\bibinfo {author} {\bibfnamefont {D.~H.}\ \bibnamefont
  {Hoe}},\ }in\ \href@noop {} {\emph {\bibinfo {booktitle} {2010 42nd
  Southeastern Symposium on System Theory (SSST)}}}\ (\bibinfo {organization}
  {IEEE},\ \bibinfo {year} {2010})\ pp.\ \bibinfo {pages}
  {258--262}\BibitemShut {NoStop}%
\end{thebibliography}
\end{document}